\pdfoutput=1
\documentclass[11pt]{article}

\usepackage[font=libertinus, citestyle=numeric]{kurbanlab}

\DeclareAffiliation{hbku}{%
  College of Science and Engineering, Hamad Bin Khalifa University, Doha, Qatar}

\DeclareAffiliation{tamu}{%
  Department of Computer and Electrical Engineering,
  Texas A\&M University, College Station, TX, USA}

\DeclareAffiliation{iub}{%
  Luddy School of Informatics, Computing, and Engineering,
  Indiana University Bloomington, Bloomington, IN, USA}

\DeclareAffiliation{tamuq}{%
  Department of Electrical and Computer Engineering,
  Texas A\&M University at Qatar, Doha, Qatar}

\DeclareAffiliation{ankara}{%
  Department of Prosthetics and Orthotics,
  Ankara University, Ankara, Turkey}

\newcommand{\tnm}[1]{\textsuperscript{#1}}

\makeatletter
\newenvironment{compute}{\KIL@back{Compute resources}}{\par\addvspace{4pt}}
\makeatother

\title{When do machine-learned exchange--correlation improvements inherit into
       density-functional tight binding?}
\RunningTitle{When do ML XC improvements inherit into DFTB?}

\Author[orcid=0000-0002-1458-302X]{Can Polat}{tamu}
\Author[corresponding=kurbanm@ankara.edu.tr, orcid=0000-0002-7263-0234]{Mustafa Kurban}{tamuq, ankara}
\Author[orcid=0000-0001-9069-770X]{Erchin Serpedin}{tamu}
\Author[corresponding=hkurban@hbku.edu.qa, orcid=0000-0003-3142-2866]{Hasan Kurban}{hbku}

\CodeURL{https://github.com/KurbanIntelligenceLab/mlxc-dftb}

\begin{document}
\maketitle

\begin{abstract}
Machine-learned exchange--correlation functionals correct band gaps at
near-semilocal cost, and density-functional tight binding reaches the
$10^3$--$10^6$-atom regime; combining them assumes a better parent yields a
better parameterization. We show it does not. Current-generation functionals are
orbital-dependent and act as generalized Kohn--Sham operators, and we prove the
parameterization channel, built on a multiplicative potential, cannot represent
them exactly. Measured by the transfer ratio, the surviving fraction of a
parent-level change, the corrections anti-transfer: coherently negative ratios
across four covalent semiconductors move the gap the wrong way; a molecular proxy
and an r$^2$SCAN control agree. The minimal-basis overgap proves to be dominated
by the on-site convention rather than by basis incompleteness, and correcting the
on-site block removes most of it; one $d$-polarization shell then closes a
further $16$ to $40\%$, a range because the placement of the empty $d$ level,
which no free-atom eigenvalue fixes, sets it. Occupied-manifold enhancements,
ionic and closed-shell repulsive potentials and rocksalt-oxide gaps inherit,
whereas elemental and III--V covalent networks inherit neither gaps nor repulsive
potentials and oxide networks inherit the latter only. We screen 23 elements and
release the parameter sets; the ratio is a cheap pre-test before any
parameterization campaign.
\end{abstract}

\printkeywords

\section{Introduction}\label{sec:intro}
Density functional theory (DFT) sets the accuracy standard for electronic-structure prediction \citep{hohenberg1964,kohn1965}, but its cost confines routine application to a few hundred atoms, well below the $10^3$--$10^6$-atom scale of extended defects, amorphous networks, and nanostructured interfaces. Density-functional tight binding (DFTB) is the physics-derived compression of DFT that reaches this regime, expanding the Kohn--Sham energy to second order in density fluctuations to obtain a semiempirical scheme roughly three orders of magnitude faster than the parent theory \citep{dftbplus2020}. Its speed rests on three approximations applied together: a minimal valence basis, a two-center expansion of the Hamiltonian and overlap into distance-dependent Slater--Koster tables \citep{slaterkoster1954}, and a local effective potential built from confined pseudo-atoms; the self-consistent-charge and third-order extensions refine the charge response on top of this fixed skeleton \citep{gaus2011dftb3}.

Machine learning has changed what can be fed into that channel. The CIDER feature set introduced nonlocal descriptors satisfying exact constraints at a cost approaching semilocal evaluation \citep{cider2022}; its successor, CIDER23X, is a nonlocal meta-generalized-gradient approximation that reproduces molecular and solid-state energetics and improves solid-state band gaps \citep{bystrom2024nonlocal}; and a further variant, CIDER24Xe, targets the derivative discontinuity directly by training on band gaps \citep{bystrom2024bandgaps}. The wider generation shares the same construction. DM21 is a local hybrid whose features include range-separated exact-exchange energy densities \citep{dm21}, and Skala is a neural functional built on the spin densities, their gradients, and the kinetic energy density \citep{skala2025}; a recent survey places all of them on the third and fourth rungs of Jacob's ladder \citep{bremond2026contemporary}. Gap-accurate exchange--correlation (XC) physics is therefore now available at a cost class compatible with the high-throughput reference calculations a parameterization campaign requires, and the obvious next move is to use it as the parent of a tight-binding model.

That move rests on an assumption that, to our knowledge, has never been measured: that a more accurate parent functional yields a better tight-binding parameterization. The assumption is load-bearing. A recent review documents the momentum behind machine-learned tight binding \citep{zou2026mltbreview}, and adaptive Slater--Koster schemes state the assumption explicitly, taking conclusions drawn at the semilocal level to carry over to higher rungs \citep{song2026adaptivesk}. If it fails, a great deal of planned work is aimed in the wrong direction.

There is a reason to doubt it that is visible before any calculation. Every functional named above is orbital-dependent: through the kinetic energy density, through exact-exchange energy densities, or through features of the density matrix. Functionals of that class are evaluated in the generalized Kohn--Sham scheme, in which the XC contribution to the effective Hamiltonian is an operator and not a multiplicative function of position \citep{kummel2008oep,perdew2017gksgaps}. The standard parameterization channel consumes a multiplicative local potential and nothing else. The mismatch is structural, it is a property of the class rather than of any one functional, and its direction is unfavorable: as machine-learned functionals climb the ladder they become harder, not easier, for the channel to represent.

Existing work approaches tight binding from the other side and leaves the question untouched. The data-driven stream fits parameters to reference data, including Hamiltonians learned from band structures or the density of states, deep-learning parameterizations, differentiable toolkits, machine-learned repulsive potentials, environment-adaptive tables, and optimizers tuned against reference bands \citep{wang2021tbhcnn,gu2024deeptb,sun2023mldftbdos,mcsloy2023tbmalt,stohr2020repulsive}. One of these is close enough to name: a machine-learning-driven DFTB scheme for the band structures of carbon, silicon, and silicon carbide \citep{fan2025mldftbbands}, the same three materials we use. That scheme fits the tables to DFT bands; we hold the tables functional-consistent and ask what the functional itself delivers, the complementary question and the one a method developer faces before any data are collected.

The parameterization tooling is mature and is not the obstacle: confined-atom generators produce Slater--Koster tables from a chosen functional \citep{koskinen2009hotbit,vandenbossche2019dftb}, and SkProgs exposes libxc, so meta-generalized-gradient and range-separated functionals are already admissible in principle \citep{skprogs,libxc2018}. What is missing is not a code path but a number: how much of a parent-level improvement is still there after the compression.

Where beyond-semilocal gap physics has entered DFTB, it has done so by extending the Hamiltonian rather than by inheriting from the functional. Long-range-corrected and hybrid DFTB add explicit nonlocal exchange terms \citep{lutsker2015lcdftb}, and dielectric-dependent hybrid schemes document the conduction-band deficiency of the minimal basis and the extended-basis remedy \citep{vanderheide2024hybriddftb}. That semilocal functionals underestimate the fundamental gap, and that the deficit is a derivative discontinuity of the exact functional rather than a mere approximation error, is long established \citep{perdewlevy1983,hybertsenlouie1986}.

Here we define the transfer ratio, the fraction of a parent-DFT property change that survives at the DFTB level when only the functional is changed, and we measure it with a potential-inversion bridge that admits any functional into a confined-atom pipeline under one overlap-certified convention. We prove that an orbital-dependent functional admits no exact multiplicative local representative, so the standard channel cannot consume the current generation of machine-learned functionals exactly. The obstruction belongs to the functional class rather than to any member of it: we prove it for the kinetic-energy-density channel, and the same non-multiplicative-operator argument carries it to the exact-exchange and density-matrix dependence of DM21 and Skala. Measured against that prediction, band-gap corrections do not merely fail to survive the compression, they reverse. Under the standard free-atom on-site convention the transfer ratio is coherently negative across diamond, silicon, 3C-SiC and boron phosphide, evaluated on one gap definition at both levels, so the compressed gap shifts opposite to the parent by nearly the full magnitude of the correction, and a molecular proxy built from the band-gap-trained CIDER24Xe agrees.

The barrier decomposes into a repairable part and a residual one. Most of the overgap of the unfitted minimal-basis tables is an artifact of the on-site convention rather than a basis deficiency; once that is corrected, a single added $d$-polarization shell closes part of what remains, by an amount that the placement of the empty $d$ level fixes rather than the physics, while the angular-channel inconsistency the theory predicts admits no repair within a shared local potential. What partially survives is specific and useful. No observable clears the transfer-ratio gate, but the occupied manifold fails less badly than the gap, and ionic and closed-shell repulsive potentials, a rocksalt-oxide gap, and same-manifold spin-flip excited-state orderings track their parent, whereas elemental and III--V covalent networks inherit neither their gaps nor their repulsive potentials and oxide networks inherit the repulsive potentials alone. The resulting map tells a practitioner where the method can be used today and what has to be built to widen it. The transfer ratio is itself the deliverable: it can be evaluated on a handful of solids before investing in a parameterization campaign, and we release three Slater--Koster parameter sets, a 23-element feasibility screen, and a single-command reproducibility suite so the pre-test has a fixed baseline.

\section{Results}\label{sec:results}

\begin{figure}[tb]
\centering
\includegraphics[width=0.90\textwidth]{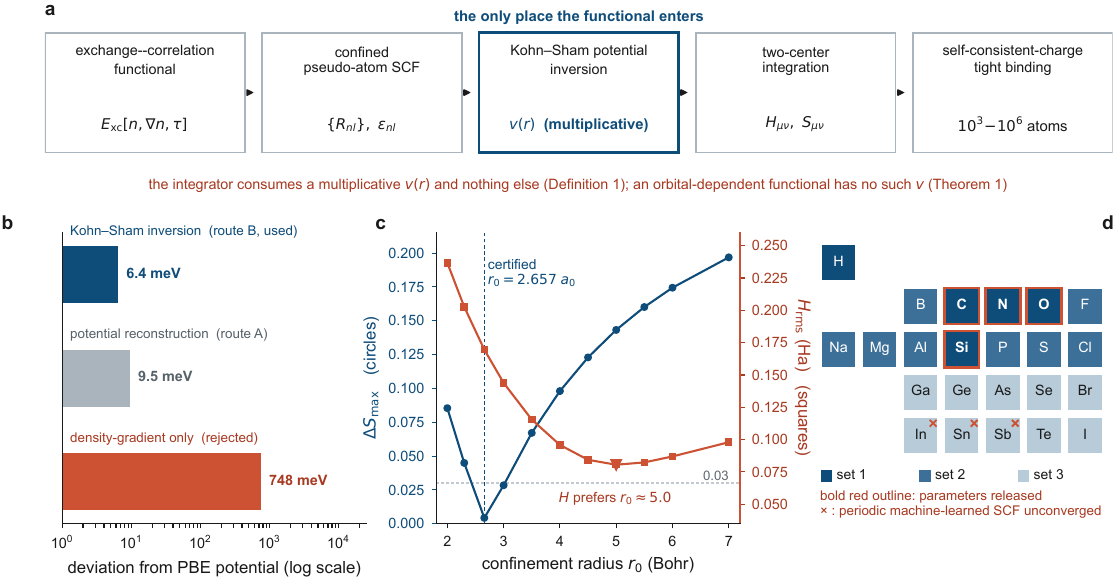}
\caption{The functional-agnostic parameterization bridge, and the three checks that certify it.
\textbf{a}, The pipeline, with what each stage produces. The functional enters at exactly one point, the Kohn--Sham potential inversion, which is also where the compression becomes lossy: the two-center integrator consumes a multiplicative $v(r)$ and nothing else, and an orbital-dependent functional admits no such $v$.
\textbf{b}, Fidelity of the inversion against a stored reference potential. The route the bridge uses and a full potential reconstruction both recover it to within a few meV; the density-gradient-only approximation fails by two orders of magnitude and was rejected.
\textbf{c}, The confinement sweep for the carbon reference. The maximum overlap deviation has a sharp minimum at the certified radius, which is what fixes the convention; the Hamiltonian root-mean-square difference does not, falling monotonically through it. The convention is certified against $S$ and against $S$ alone.
\textbf{d}, Coverage. Elements passing the atomic feasibility gates across three element sets; the periodic machine-learned self-consistent field is the binding constraint.
}
\label{fig:bridge}
\end{figure}

\subsection{A functional-agnostic parameterization bridge}\label{sec:r1}

A single functional-agnostic bridge lets one parameterization pipeline accept any XC functional and hold everything else fixed, so the functional becomes the only varied ingredient. We parameterize density-functional tight binding from a chosen XC functional through a fixed sequence: a confined pseudo-atom self-consistent field supplies the compressed valence orbitals and the local reference potential; a Kohn--Sham potential inversion converts the confined solution of any functional into an effective potential that hotcent integrates into signed two-center Hamiltonian and overlap tables; and those tables, with a fractional-occupation Hubbard $U$ and spin constants, close the self-consistent-charge model (\cref{fig:bridge}a, stages~1--4). The inversion step is what makes the pipeline functional-agnostic. The two-center integrator consumes a multiplicative local potential $v(r)$ and reads its XC contribution from libxc \citep{libxc2018}, so a machine-learned functional, which is not in libxc and whose XC operator is not multiplicative, cannot enter through a closed-form energy expression. It enters instead through the best local potential its converged confined-atom solution admits.

The instrument we carry through the whole study is the transfer ratio: the change a tight-binding property inherits when only the parent functional is swapped, divided by the change the swap produces at the parent-DFT level, so that one denotes full inheritance and zero none. We intend it as a pre-test any group can run on a handful of solids before committing to a parameterization campaign, and the theory of \cref{sec:r3} is the warrant for reading a sub-threshold ratio as a structural verdict rather than a numerical accident.

The comparison rests on two certifications. The overlap tables reproduce published reference parameterizations: across all fifteen element pairs of the H/C/N/O/Si base set the maximum overlap deviation from the reference Slater--Koster tables stays within a few percent (per-pair breakdown in \ref{sn:overlap}), with two elements taking element-specific radii and silicon needing an explicit $3d$ polarization channel for the O--Si pair (\cref{fig:bridge}c). The confinement radius is set once by this overlap criterion and held identical across all functionals, so that any downstream difference reflects the functional and not a re-tuned basis. The certification is on the overlap and not on the Hamiltonian: at the certified radius the Hamiltonian root-mean-square difference from the reference tables is substantial and is minimized at a different radius altogether. An unfitted convention has no reason to reproduce the reference Hamiltonian, and this discrepancy is the origin of the overgap analysed below.

The inversion also reproduces a known potential: applied to a PBE confined atom whose effective potential is stored exactly, the Kohn--Sham inversion the bridge uses recovers that potential to a median of $6.4$~meV and the injected tables match the native ones bit for bit; a full potential-reconstruction route reaches $9.5$~meV, while the density-gradient-only approximation fails at $748$~meV (\cref{fig:bridge}b; \ref{sn:bridge}). The signed C/N/O/Si tables are complete, and the resulting two-center integrals carry the functional's imprint element by element: for silicon the $\mathrm{sp}\sigma$ channel shifts most strongly between the machine-learned and semilocal functionals while the $\mathrm{pp}\sigma$, $\mathrm{pp}\pi$, and $\mathrm{ss}\sigma$ channels move little. The fractional-occupation Hubbard $U$ is smooth for all three functionals tested (\ref{sn:atomic}).

\subsection{Band-gap corrections do not survive the compression}\label{sec:r2}

With the bridge fixed, the parent-level band-gap correction does not survive the compression: under the physical free-atom on-site convention the transfer ratio is coherently negative, so a more accurate parent buys a tight-binding gap that moves the wrong way. We evaluate the ratio for the fundamental gap and the occupied bandwidth, across the covalent semiconductors diamond, silicon, 3C-SiC and boron phosphide (and one ionic rocksalt, MgO), with a decision threshold of $0.5$ marking useful transfer. At the parent level the machine-learned functional opens the gap uniformly and substantially, by $+2.1$ to $+3.6$~eV across the four covalent solids. A failure downstream is therefore a property of the compression rather than of the functional, provided the compression is otherwise correctly executed. The direction of this parent shift is toward experiment (all underestimated by PBE and moved upward by the machine-learned functional), so it is a correction rather than an arbitrary change; the experimental references are plotted at the indirect definition in \cref{fig:bands}d--f, and the parent gaps are converged at the production settings (\ref{sn:conv}).

\begin{figure}[tb]
\centering
\includegraphics[width=0.90\textwidth]{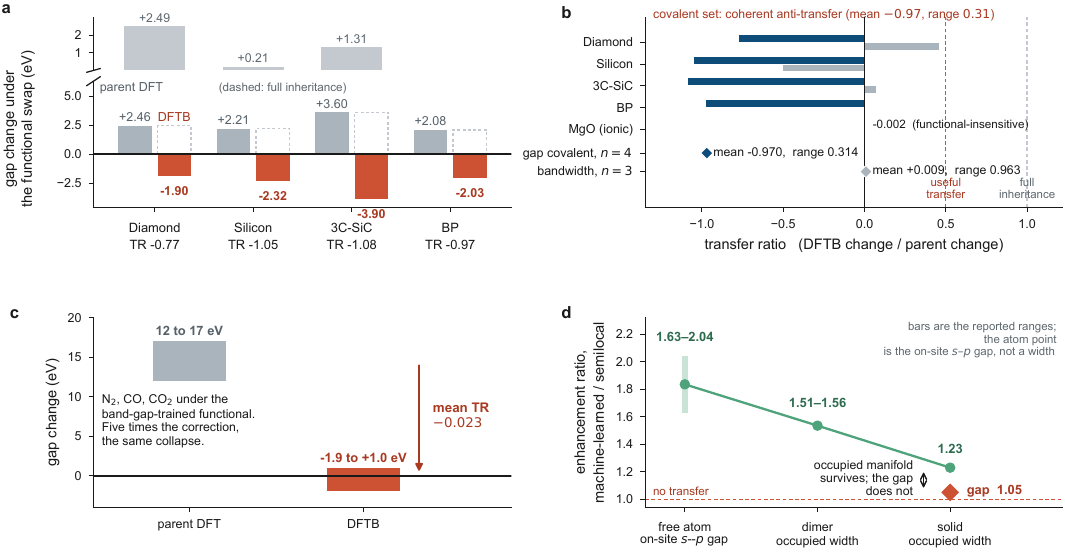}
\caption{The band-gap correction reverses under the compression, and what survives instead.
\textbf{a}, Input against output: the parent gap opens (grey; dashed outlines mark full inheritance) while the compressed DFTB gap closes (red). The upper segment carries the scale this is read against, the residual overgap of the same tables under the same convention: $+2.49$, $+0.21$ and $+1.31$~eV for diamond, silicon and 3C-SiC.
\textbf{b}, The same result as a ratio, against the $0.5$ useful-transfer threshold. Gap ratios for the covalent solids, the near-zero ionic case (MgO), and the bandwidth ratios (a width quantity, convention-independent) reported for the three solids that carry them; none clears the gate.
\textbf{c}, The molecular proxy on N$_2$, CO and CO$_2$, with a parent correction five times larger. Parent and tight-binding responses are drawn as ranges.
\textbf{d}, What survives. The free-atom on-site $s$--$p$ opening persists as an occupied-width enhancement into the dimer and solid while the solid gap ratio barely moves, the partition the theory predicts.
}
\label{fig:gate}
\end{figure}

The measurement bears this out (\cref{tab:tr}, \cref{fig:gate}a,b). Under the free-atom convention the failure is sharper than a null result: the correction \emph{anti-transfers}. The transfer ratio is negative and large for every covalent solid, and tightly clustered: the range across the four covalent semiconductors is a small fraction of the mean magnitude. The anti-transfer is therefore coherent across the covalent set, not the sign-incoherent scatter around zero of the confined-convention analysis. The one ionic solid, magnesium oxide, behaves differently: its free-atom ratio is near zero, essentially no transfer of the parent correction, because the rocksalt gap is set by the cation--anion on-site offset and is nearly insensitive to the functional.

A molecular proxy built from the flagship band-gap-trained functional on N$_2$, CO, and CO$_2$ gives an average confined-convention transfer ratio of $-0.023$ (\cref{fig:gate}c), so the failure is not specific to one functional. The proxy also separates the functional from the representation: the parent-level corrections are large while the tight-binding response is small and of the wrong sign. The functional produces the correction; the representation has no channel to carry its orbital-dependent part, and what little it carries has the wrong sign.

What survives best is an occupied-manifold signature that dilutes with atomic coordination, measured as an \emph{enhancement} ratio: the CIDER-to-PBE occupied width, unnormalized by the parent-level change and therefore distinct from the transfer ratio. Measured by the transfer ratio, the occupied bandwidth also fails the $0.5$ gate, at a mean of $+0.009$ with a range of $0.963$ across the three solids that report it; it fails less badly than the gap but far less coherently, the wide range reflecting that one solid inherits part of the width change while another reverses it. The functional's on-site $s$--$p$ gap opening, roughly a factor of two at the free atom, survives as an occupied-width enhancement at the dimer, weakens in the solid, and all but vanishes in the solid gap (\cref{fig:gate}d): the signature dilutes as coordination grows. The surviving signature is size-stable and cheap: a thousand-atom silicon supercell converges in seconds on a single CPU, at essentially the semilocal-tables cost, with a size-independent gap, so the transferable part of the functional's effect is deployable at the parent's cost.

\begin{table}[tb]
\caption{The band-gap correction does not merely fail to survive, it moves the tight-binding gap the wrong way. All free-atom entries are evaluated on the fundamental band-path gap at both levels. The magnesium-oxide gap change is of order $0.01$~eV, so the ionic case is functional-insensitive rather than positively transferring. A ratio of one is full inheritance and $0.5$ the useful-transfer threshold; the confined ratios, which the on-site convention error drove toward zero, are given for comparison in the last column.}\label{tab:tr}%
\begin{tabular*}{\textwidth}{@{\extracolsep{\fill}} l
  S[table-format=+1.3] S[table-format=+1.3]
  S[table-format=+1.3] S[table-format=+1.3] @{}}
\toprule
 & {Parent DFT} & {DFTB (free-atom)} & \multicolumn{2}{c}{Gap transfer ratio} \\
\cmidrule{4-5}
System & {$\Delta E_{\mathrm{g}}$ (eV)} & {$\Delta E_{\mathrm{g}}$ (eV)\tnm{1}}
       & {Free-atom} & {Confined\tnm{2}} \\
\midrule
\multicolumn{5}{@{}l}{\itshape Covalent semiconductors} \\
Diamond & +2.459 & -1.896 & -0.771 & +0.128 \\
Silicon & +2.211 & -2.323 & -1.051 & -0.325 \\
3C-SiC  & +3.595 & -3.900 & -1.085 & -0.058 \\
BP      & +2.082 & -2.029 & -0.974 & -0.024 \\
\quad Mean, $n=4$  & {--} & {--} & -0.970 & -0.070 \\
\quad Range & {--} & {--} &  0.314 &  0.453 \\
\addlinespace
\multicolumn{5}{@{}l}{\itshape Ionic (rocksalt)\tnm{3}} \\
MgO     & +6.100 & -0.010 & -0.002 & -0.174 \\
\addlinespace
\multicolumn{5}{@{}l}{\itshape Molecular proxy, $n=3$, a different functional\tnm{4}} \\
\quad Mean  & {$+12$ to $+17$} & {$-1.9$ to $+1.0$} & {--} & -0.023 \\
\bottomrule
\end{tabular*}
\tabnote{\textsuperscript{1}\,Free-atom-convention DFTB gap change $\Delta E_{\mathrm{g}}^{\mathrm{DFTB}} = E_{\mathrm{g}}^{\mathrm{CIDER}} - E_{\mathrm{g}}^{\mathrm{PBE}}$, from the released tables (CIDER: diamond $5.793$, Si $0.156$, 3C-SiC $2.915$~eV; PBE: $7.689$, $2.479$, $6.814$~eV; BP $3.418$ against $5.447$, MgO $8.439$ against $8.450$~eV). The free-atom transfer ratio is this divided by the parent $\Delta E_{\mathrm{g}}$. Both sides of every ratio use the same gap definition and the same $k$-sampling protocol, given in Methods; for diamond and silicon the fundamental and $\Gamma$-point definitions coincide, and for 3C-SiC the fundamental gap lies $0.03$~eV below the $\Gamma$ gap. \par\smallskip \textsuperscript{2}\,Confined (as-released) on-site convention; the on-site inflation drives these toward zero. The corresponding bandwidth changes are in Supplementary \cref{tab:bw}. \par\smallskip \textsuperscript{3}\,Per-solid provenance and pipeline settings for the magnesium-oxide and boron-phosphide pairs are in \ref{sn:r19}. \par\smallskip \textsuperscript{4}\,N$_2$, CO and CO$_2$ under the density-matrix, band-gap-trained functional, which has no periodic route (confined convention). The parent correction is five times larger and the collapse is the same.}
\end{table}

\subsection{Theory: what the parameterization channel can carry}\label{sec:r3}

Two statements must be kept apart in what follows. The first is a theorem: an orbital-dependent functional has no exact multiplicative local representative, so the channel cannot consume it exactly. The second is the measurement above: the compressed gap moves opposite to the parent. The theorem identifies the ingredient that opens the parent gap as one the channel cannot carry, and so identifies the mechanism at risk; the conditions under which it also accounts for the measured anti-transfer are stated with the propositions below.

Both machine-learned functionals used here are nonlocal meta-generalized-gradient approximations: their exchange energy density depends on the kinetic energy density $\tau(\mathbf{r}) = \tfrac{1}{2}\sum_i^{\mathrm{occ}} |\nabla\psi_i(\mathbf{r})|^2$, and CIDER24Xe additionally on features of the density matrix. They are not special in this. DM21 carries range-separated exact-exchange energy densities and Skala carries $\tau$, so the current generation shares the property that matters here. Write
\begin{equation}\label{eq:f}
f(\mathbf{r}) \;:=\; \frac{\partial e_{\mathrm{xc}}}{\partial \tau}(\mathbf{r}),
\end{equation}
a dimensionless field that vanishes identically for a local or generalized-gradient functional and does not vanish for anything above the second rung. Differentiating $E_{\mathrm{xc}}[n,\nabla n,\tau]$ with respect to $\psi_i^{*}$, with $f$ as defined in \cref{eq:f}, gives the generalized Kohn--Sham equation
\begin{equation}\label{eq:gks}
-\tfrac{1}{2}\,\nabla\!\cdot\!\big[(1+f(\mathbf{r}))\,\nabla\psi_i\big] \;+\; v(\mathbf{r})\,\psi_i \;=\; \varepsilon_i\,\psi_i ,
\end{equation}
in which $v$ collects every multiplicative contribution (external, Hartree, confinement, and the $n$- and $\nabla n$-derived part of $e_{\mathrm{xc}}$) and the $f$ term is an operator, not a function of position. \cref{eq:gks} is standard; the two definitions that follow are what let us use it.

\begin{definition}[The parameterization channel]\label{def:channel}
The standard channel builds its two-center tables as $H_{\mu\nu} = \langle \phi_\mu | -\tfrac{1}{2}\nabla^2 + v_{\mathrm{sup}} | \phi_\nu\rangle$ and $S_{\mu\nu} = \langle \phi_\mu | \phi_\nu\rangle$, where $\{\phi_\mu\}$ are confined atomic orbitals and $v_{\mathrm{sup}}$ is a superposition of multiplicative atomic potentials. The channel can express the free-particle kinetic operator and a multiplicative potential, and nothing else.
\end{definition}

\begin{definition}[Channel representability]\label{def:rep}
An XC functional is \emph{channel-representable} at a given confined atom if there exists a single multiplicative $v_{\mathrm{ch}}(r)$ whose radial Kohn--Sham equation admits every one of the functional's converged confined-atom radial functions $R_{nl}$ as a solution at the functional's own eigenvalue $\varepsilon_{nl}$.
\end{definition}

Channel representability is exactly the property the channel needs and never checks. It is decidable, and the following identity decides it.

\begin{proposition}[The $l$-resolved local representative]\label{prop:vl}
Let $R_{nl}$ solve the radial reduction of the generalized Kohn--Sham equation under Assumptions~A1--A3 of \ref{app:assump}. Then the unique multiplicative potential reproducing $R_{nl}$ at $\varepsilon_{nl}$ in an ordinary radial Kohn--Sham equation is
\begin{equation}\label{eq:vl}
v_l(r) \;=\; v(r) \;+\; f(r)\,\frac{l(l+1)}{2r^{2}} \;-\; \frac{1}{2r^{2}R_{nl}(r)}\,\frac{\mathrm{d}}{\mathrm{d}r}\!\left[r^{2} f(r)\, R_{nl}'(r)\right].
\end{equation}
\end{proposition}

\noindent Proof in \ref{app:vl}. The two $f$-dependent terms of \cref{eq:vl} both carry $l$, the first algebraically through the centrifugal term and the second through the orbital, and they are not independent. Extracting the net $l$-dependence requires a further step, given in \ref{app:dvdl}: the $l(l+1)$ terms cancel identically, so the local representative acquires no quadratic $l$-weighting and $v_l - v$ is exactly affine in $l$ at fixed reduced function, with every surviving $l$-dependent term proportional to $f$ or to $f'$. The difference between two channels does not follow from that structure alone; the nucleus settles it, since the relative behaviour of the reduced functions there is fixed by the cusp condition.

\begin{theorem}[No exact local representative]\label{prop:nonmult}
Let $Z$ be the nuclear charge and impose the cusp condition $S_{nl}'(0)/S_{nl}(0) = -Z/(l+1)$. Then, as $r\to 0$,
\begin{equation}\label{eq:cusp}
v_l(r) - v(r) \;=\; \frac{l}{2(l+1)}\,\frac{2Z f(0) - (l+1) f'(0)}{r} \;+\; O(1),
\end{equation}
so that
\begin{equation}\label{eq:vpvs}
v_p(r) - v_s(r) \;=\; \frac{Z f(0) - f'(0)}{2\,r} \;+\; O(1).
\end{equation}
If $f\equiv 0$ the coefficient vanishes, $v_l = v$ for every $l$, and the functional is channel-representable with $v_{\mathrm{ch}} = v$. If $f\not\equiv 0$ and $f'(0)\neq Z f(0)$, the $s$ and $p$ representatives differ by a term that diverges at the nucleus, no multiplicative $v_{\mathrm{ch}}$ can equal both, and the functional is not channel-representable. The excluded case, $f'(0) = Z f(0)$, is a single scalar coincidence between the functional and the nucleus it is evaluated at, and no functional in use satisfies it identically across the periodic table.
\end{theorem}

\begin{figure}[tb]
\centering
\includegraphics[width=0.90\textwidth]{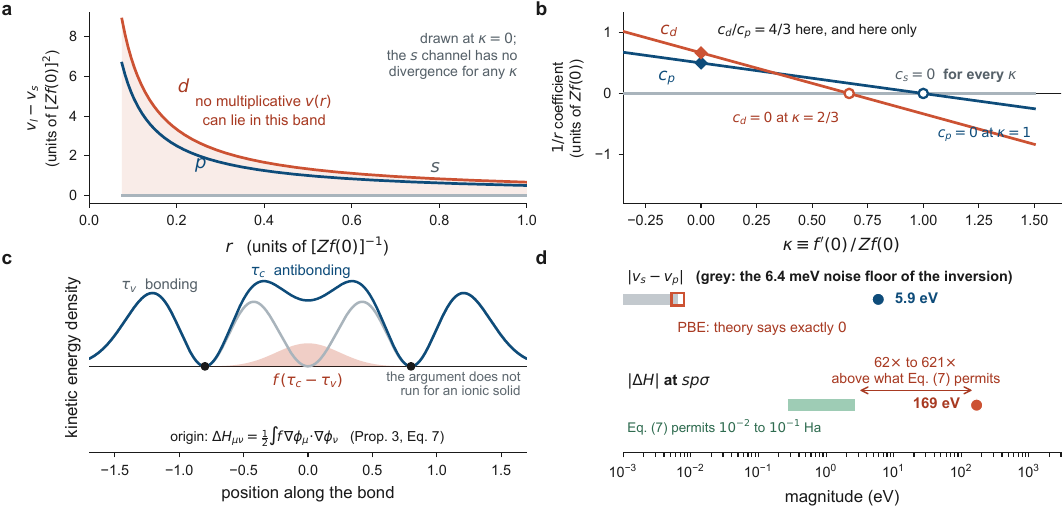}
\caption{Why the parameterization channel cannot carry an orbital-dependent functional, and the two measurements that decide whether it does. $\kappa \equiv f'(0)/Z f(0)$ is the dimensionless number the functional fixes at the nucleus.
\textbf{a}, The $l$-resolved local representative with the cusp condition imposed, at $\kappa=0$. Each angular channel sees a different potential and the difference diverges as $1/r$, so no multiplicative $v(r)$ lies in the shaded band; only the $s$ divergence $c_s=0$ is parameter-free.
\textbf{b}, The cusp coefficients against $\kappa$. Both are linear and vanish at different $\kappa$ ($1$ and $2/3$), so they never vanish together; the ratio $c_d/c_p=4/3$ at $\kappa=0$ only.
\textbf{c}, Where the discarded gap-shift term acts. Sketched for a homonuclear pair; the argument does not run for ionic solids.
\textbf{d}, Two measurements on a common scale. At the production inversion's matching radius the channel splitting $|v_s-v_p|$ is $5.92$~eV, three orders above that inversion's $6.4$~meV noise floor; under PBE the theory predicts exactly zero and the negative control confirms it at the floor.
}
\label{fig:theory}
\end{figure}

\noindent Proof in \ref{app:thm}. The one degenerate case closes once a third channel is present, since no single $\kappa \equiv f'(0)/Zf(0)$ removes both the $p$- and the $d$-channel divergence (\cref{fig:theory}b; coefficients and argument in \ref{app:kappa}). The conclusion constrains the functional class rather than any member of it, since $f\not\equiv 0$ is the only condition it uses, so exchanging one machine-learned functional for another does not lift it.

The error the channel makes is not only nonzero but available in closed form.

\begin{proposition}[The term the channel discards]\label{prop:dH}
Under Assumptions~A1--A4, the exact generalized Kohn--Sham matrix element differs from the channel's by
\begin{equation}\label{eq:dH}
\Delta H_{\mu\nu} \;=\; H^{\mathrm{GKS}}_{\mu\nu} - H^{\mathrm{ch}}_{\mu\nu} \;=\; \frac{1}{2}\int f(\mathbf{r})\; \nabla\phi_\mu(\mathbf{r})\cdot\nabla\phi_\nu(\mathbf{r})\;\mathrm{d}^{3}r ,
\end{equation}
a gradient overlap weighted by $f$. For non-constant $f$ there is no multiplicative $w$ with $\Delta H_{\mu\nu} = \langle \phi_\mu | w | \phi_\nu\rangle$, so no confinement radius, superposition convention, or inversion route removes it. It is the exact residual of the compression.
\end{proposition}

\noindent Proof in \ref{app:dH}. \cref{eq:dH} says what is lost. The next result says where the loss lands, and states the condition under which it explains our measurement.

\begin{proposition}[Where the loss lands]\label{prop:gap}
Let $\chi_v$ and $\chi_c$ be the valence- and conduction-edge eigenstates of the solid's channel Hamiltonian, distinct from the confined-atom orbitals $\psi_{nl}$ and from the basis functions $\phi_\mu$, and let $\tau_k = \tfrac{1}{2}|\nabla\chi_k|^{2}$ be the kinetic energy density of the single state $\chi_k$. To first order in $\Delta H$,
\begin{equation}\label{eq:dEg}
\delta\varepsilon_k \;=\; \int f(\mathbf{r})\,\tau_k(\mathbf{r})\,\mathrm{d}^{3}r ,
\qquad
\delta E_{\mathrm{g}} \;=\; \int f(\mathbf{r})\,\big[\tau_c(\mathbf{r}) - \tau_v(\mathbf{r})\big]\,\mathrm{d}^{3}r .
\end{equation}
An antibonding conduction-edge state changes sign between the atoms and therefore carries more curvature in the bonding region than the bonding valence-edge state it is paired with, so $\tau_c > \tau_v$ there and $\delta E_{\mathrm{g}}$ does not vanish (\cref{fig:theory}c). The argument is physical rather than a proof, and \ref{app:gap} says where it can fail.
\end{proposition}

\noindent What \cref{prop:gap} establishes is that a gap correction of the form of \cref{eq:dEg} is discarded by the channel. What it does \emph{not} establish is that the parent's \emph{entire} gap correction is of that form. The parent's correction has two parts, one carried by the multiplicative $v$ and one by $f$, and the channel can express the first. The measured anti-transfer therefore follows from the theory only if the $f$ channel dominates the parent's gap correction. That condition is quantitative and evaluable from parent-level quantities alone; we defer it to future work, and until it is tested this proposition explains why the gap is \emph{at risk} rather than why the measured transfer ratio reverses.

\begin{corollary}[What can transfer]\label{cor:transfer}
The channel can express any multiplicative potential exactly, so the part of a functional's effect that acts through $v$, which acts on the density, is in principle inheritable. The part that acts through $f$, which acts on orbital curvature, is not inheritable under any choice of $v$. Occupied-manifold widths and on-site $s$--$p$ splittings are of the first kind and the fundamental gap is of the second.
\end{corollary}

\noindent The qualification in \cref{cor:transfer} is needed because the production bridge inverts a converged solution and, by the theorem, returns some $v_l$ rather than $v$ itself. The partition it predicts nonetheless matches what the measurements above deliver.

The theory makes two predictions this study can test, set out in Supplementary \cref{tab:pred} with the observation that would have refuted each, and evaluated in turn below.

The near-nucleus law of P2 ties a measurable number to the functional rather than to the inversion: \cref{eq:cusp} fixes $c_s = 0$ identically for any functional, makes $c_p = \tfrac{1}{2}(1-\kappa)Zf(0)$ and $c_d = \tfrac{1}{3}(2-3\kappa)Zf(0)$ linear in the single number $\kappa = f'(0)/Zf(0)$ computable from the functional on its own converged density, and sets their ratio to $4/3$ at $\kappa = 0$ and nowhere else, the value at which the $l/(l+1)$ weights of \cref{fig:theory}a are drawn, excluding outright the naive $l(l+1)$ reading that would force the ratio to $3$ regardless of $\kappa$.

Evaluated on their own converged confined densities, the machine-learned functional (CIDER24Xe) gives $Zf(0)\approx0.070$~Ha, near-constant across carbon, nitrogen and oxygen, with $\kappa\approx6.0$--$6.3$, far from both degenerate values ($1$ and $2/3$), so $c_p$ and $c_d$ are both substantial. This magnitude is strongly functional-dependent: the lower-rung CIDER23X gives $Zf(0)\approx10^{-3}$~Ha, two orders of magnitude smaller. Since $f\not\equiv0$ for both, the theorem applies to both regardless; what varies with rung is only the size of this particular near-nucleus term. Silicon's $f'(0)$ is not robustly resolved on the present radial grid and its $\kappa$ is reported as indeterminate. This is the P2 measurement recorded in Supplementary \cref{tab:pred}.

The P1 negative control of Supplementary \cref{tab:pred} settles the magnitude question. For PBE $f\equiv0$, so the theorem requires the two-channel inverted potentials to coincide exactly; the measured $|v_s-v_p|$ has a \emph{median} of $3.85$~meV (carbon) and $2.72$~meV (silicon), at or below the $6.4$~meV inversion noise floor, confirming the inversion is orbital-independent for PBE. The maximum pointwise residual is large but localized at the radial node, where the inversion is ill-conditioned, so a fixed-radius extraction reports node contamination as a pointwise extremum rather than a physical hopping integral; the robust median vanishes as the theory predicts and the closed-form bound is not contradicted.

The residual the channel discards is visible in the inversion itself (\cref{fig:theory}d), and it is best read from the channel potentials rather than from any single two-center integral. On the node-masked radial grid, inverting the confined $s$ and $p$ solutions separately yields local potentials whose asymmetry $|v_s-v_p|$ has a median of $1.4$~Ha for carbon, $2.0$~Ha for nitrogen, $1.4$--$1.5$~Ha for oxygen and $3.2$~Ha for silicon, against $0.006$--$0.061$~Ha for the PBE control, whose $f\equiv0$ makes the theorem require coincidence: a factor of $24$ to $318$ above the semilocal floor. Both machine-learned functionals agree element by element to within $6\%$ despite near-nucleus coefficients differing by two orders of magnitude (Supplementary \cref{tab:onsite}), so the $l$-channel asymmetry is carried by the orbital dependence itself and does not track the functional's rung, unlike the near-nucleus term of \cref{eq:cusp}. The robust statement is the channel-potential asymmetry: a single local potential cannot simultaneously serve the $s$ and $p$ channels the functional distinguishes, and the gap-setting behaviour is the one that breaks. The median is the reported statistic because the pointwise extrema are artifacts of a fixed-radius extraction in the node region rather than physical hopping integrals.

The wall is not a silicon pathology: the asymmetry is present for every atom tested and is a property of orbital dependence ($f\not\equiv0$) rather than of any one element, though its severity is element-dependent (Supplementary \cref{tab:e2}).

\subsection{The overgap of the unfitted minimal basis}\label{sec:r3b}

\begin{figure}[tb]
\centering
\includegraphics[width=0.90\textwidth]{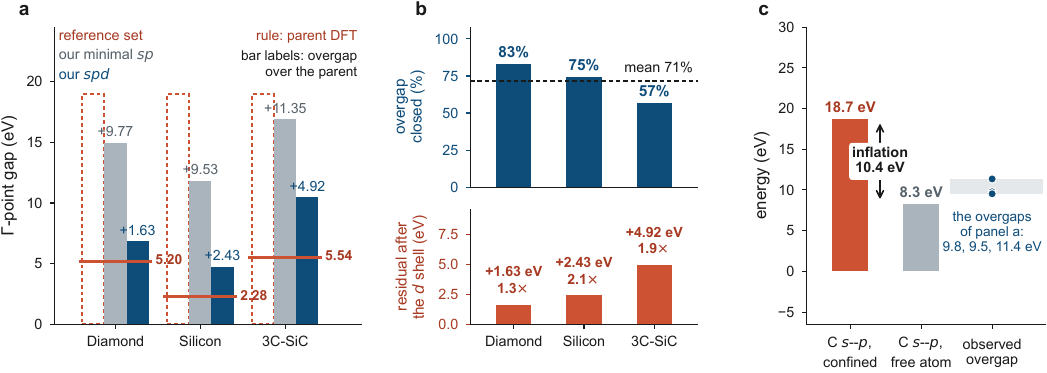}
\caption{The overgap of the unfitted tables, what a single $d$ shell removes of it, and the two things that would undo the claim.
\textbf{a}, $\Gamma$-point gaps; red rules mark the parent (identical settings). The minimal $sp$ tables overgap the parent by $\sim\!10$~eV; adding a $d$-polarized $spd$ set (confinement radii and on-site eigenvalues held fixed) reduces the overgap several-fold (per-solid values in \cref{tab:spd}). The published-parameter baseline is given in \cref{fig:bands}; under the free-atom convention the excess is shared with fitted sets, not a property of our convention alone.
\textbf{b}, The closure and its residual, drawn at the confined on-site convention, which isolates the basis-completeness effect; under the free-atom convention adopted in this work the closure is smaller and depends on the placement of the empty $d$ level, as \cref{fig:dconv} shows. The residual grows with polarity and is intrinsic to the two-center on-site approximation, not removed by further basis enlargement (per-solid closures and residuals in \cref{tab:spd}).
\textbf{c}, The on-site convention, tested and corrected. The standard scheme puts \emph{free}-atom eigenvalues on the diagonal $H^{0}_{\mu\mu}$; the confined convention places \emph{confined} eigenvalues there instead (every on-site $\varepsilon_p>0$). Rebuilding with free-atom eigenvalues collapses the $sp$ overgap, confirming the convention was the dominant contributor to panel \textbf{a}.
}
\label{fig:basis}
\end{figure}

A second, independent barrier operates at the level of the basis, and unlike the first it is partly repairable: the unfitted minimal basis inflates the gap far beyond the correction, and almost all of that inflation turns out to be a bookkeeping convention rather than a deficiency of the basis. Our minimal single-$s$--$p$ silicon parameterization places the $\Gamma$-point gap at $11.81$~eV, against a parent direct gap of $2.28$~eV (\cref{fig:basis}a). Attributing this overgap to the minimal basis requires the on-site convention to be stated. In the standard scheme the diagonal elements $H^{0}_{\mu\mu}$ are the \emph{free}-atom eigenvalues, with the confined orbitals entering only the two-center integrals; the confined \emph{carbon} atom, taken as the illustrative case, instead has $\varepsilon_s = -0.162$~Ha, $\varepsilon_p = +0.525$~Ha (an $s$--$p$ splitting of $18.7$~eV against a free-atom value near $8.3$~eV, and a positive occupied-valence eigenvalue). Whether those confined values occupy the on-site block of the tables is therefore load-bearing for the overgap and for the transfer ratio, and we tested it directly.

The test is decisive. Under the confined convention every homonuclear on-site $\varepsilon_p$ is positive ($+4.84$ to $+16.36$~eV across all sets, impossible for a bound valence $p$ level). Recomputing free-atom valence eigenvalues with the same solvers and placing them in every homonuclear on-site block, holding the Hubbard $U$, occupations, and all two-center tables fixed, collapses the $\Gamma$-gap by $9.5$ to $15.4$~eV, bringing diamond and 3C-SiC close to experiment and overcorrecting silicon, a minimal-$sp$ limitation. The periodic overgap is therefore overwhelmingly the confined on-site convention, not functional physics (\cref{fig:basis}c). The free-atom tables are the ones carried through every result in this work and the ones released (\ref{sn:onsite}).

The overgap itself is a property of tables built under a fixed, overlap-certified confinement convention and never refitted to band structures; production sets tune a separate density compression radius against the parent bands and do not show an excess of this size. The transfer ratio is nevertheless a difference of differences taken within one convention. The correction the functional carries is an order of magnitude smaller than the overgap of the representation into which it is compressed, so within this convention even a faithfully inherited correction would be masked.

\begin{figure}[tb]
\centering
\includegraphics[width=0.90\textwidth]{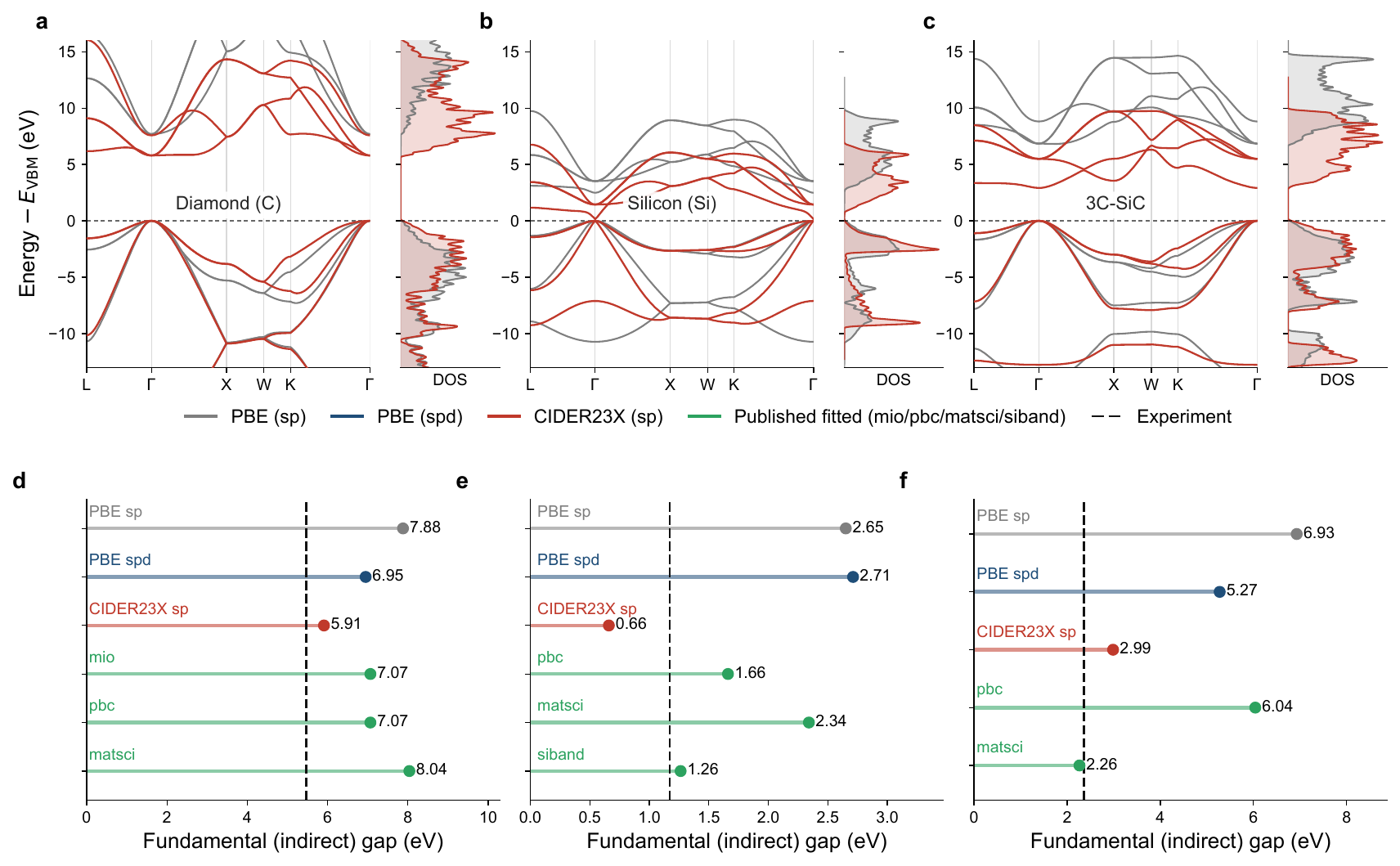}
\caption{Band structures, densities of states, and the fitted-table baseline, in the free-atom on-site convention. \textbf{a--c}, Band structures along L--$\Gamma$--X--W--K--$\Gamma$ with the total density of states, for diamond, silicon, and 3C-SiC under the semilocal reference (PBE, $sp$ and $spd$) and the machine-learned functional (CIDER23X, $sp$); energies referenced to the valence-band maximum. \textbf{d--f}, Fundamental gaps of each functional-consistent table and of the fitted published sets, run in the same DFTB+ binary, against the experimental gap (dashed).
}
\label{fig:bands}
\end{figure}

Fitted reference sets (mio-1-1, pbc-0-3, matsci-0-3, siband-1-1) run in the same DFTB$+$ build alongside the free-atom PBE parameterization complete this comparison. The fitted-vs-experiment agreement is per-solid: fitted tables land near experiment for silicon and 3C-SiC, but for diamond every fitted table overgaps, \emph{above} the functional-consistent CIDER23X gap, which is the closest of all diamond entries to the experimental $5.47$~eV (\cref{fig:bands}d--f). Minimal-basis DFTB gaps are thus dominated by table construction, not by the functional, and the functional-consistent tables are competitive with fitted ones rather than anomalous.

How much of what remains is basis rather than physics is measurable, and the answer depends on the on-site convention in a way that is itself the result. Rebuilding the carbon and silicon electronic tables with a single added $d$-polarization shell, holding the confinement radii and on-site eigenvalues fixed so that only the valence $l$-set changes, isolates the basis-completeness effect. Held at the confined convention on both sides, where the $sp$ and $spd$ sets differ in nothing but the valence $l$-set, the added shell removes a mean $71\%$ of the overgap for every solid tested (\cref{fig:basis}b and \cref{tab:spd}). Under the convention adopted throughout this work the picture changes, because the free-atom correction has already removed most of the overgap by itself: the $sp$ excess falls from roughly $10$~eV to between $0.2$ and $2.5$~eV, leaving little for a polarization shell to close. Recomputing the same decomposition with free-atom $s$ and $p$ on-site energies gives a mean closure of $16.3\%$ on the fundamental definition and $1.0\%$ at $\Gamma$, with silicon turning slightly negative in both. The residual is therefore small in absolute terms rather than large and mostly repairable.

\begin{figure}[tb]
\centering
\includegraphics[width=0.90\textwidth]{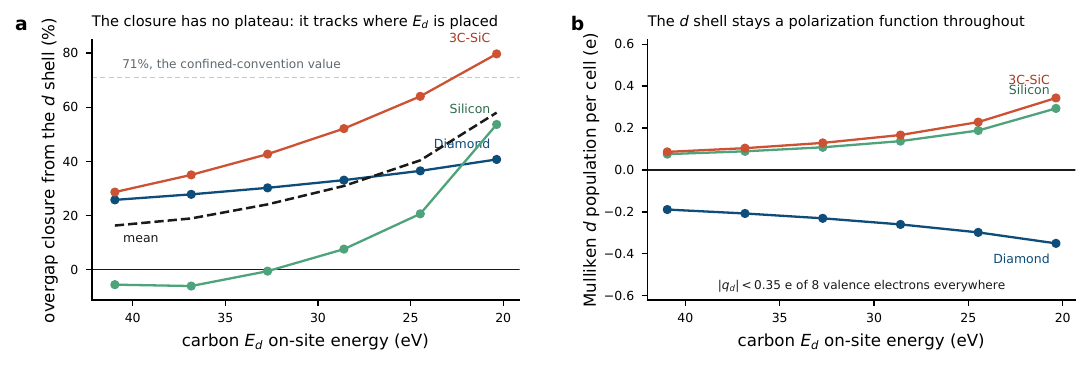}
\caption{Under the convention the overgap closure tracks the placement of the empty $d$ level, which no free-atom eigenvalue fixes.
\textbf{a}, Mean and per-solid closure as the carbon $E_d$ is swept from the confined value ($+40.96$~eV) toward the valence shell, with silicon shifted in proportion. The dashed rule marks the closure obtained when the confined convention is used on both the $sp$ and $spd$ sides.
\textbf{b}, The Mulliken $d$ population across the same sweep, against the eight valence electrons of the cell; placements further down, at the free-atom $p$ level, take the $d$ manifold to a valence-shell population and are excluded on that basis. All gaps are fundamental band-path gaps; $n=3$ solids, confinement radii and free-atom $s$ and $p$ on-site energies held fixed throughout.}
\label{fig:dconv}
\end{figure}

The remaining closure depends on the convention adopted for the empty $d$ channel, for which no bound free-atom level exists, and that dependence is the result rather than a nuisance to be resolved by a choice. Sweeping the carbon $E_d$ from the confined value of $+40.96$~eV down to $+20.36$~eV, with silicon shifted in proportion, moves the mean closure smoothly from $16.3$ to $58.0\%$ with no plateau anywhere along the range (\cref{fig:dconv}a). Two placements are physically distinguishable within that range: retaining the confined $E_d$, which is what the $spd$ tables as generated contain, and shifting $E_d$ rigidly by the amount the valence $p$ level moves under the free-atom correction, which preserves the $d$-to-$p$ separation of the generating atom and gives $40.4\%$. Placing the empty $d$ at the free-atom $p$ level is excluded: the $d$ manifold then falls below the conduction edge, takes up $19$ to $87\%$ of the valence population, and returns negative gaps for all three solids, so it is no longer a polarization function. Over the entire admissible range the $d$ population stays below $0.35$~e of eight valence electrons (\cref{fig:dconv}b), and the closure never reaches the value obtained under the confined convention.

The $d$ shell reduces the residual overgap for diamond and 3C-SiC at every placement tested, it never removes it, and the residual never falls below roughly one eV. For silicon the sign is not even stable, the closure passing through zero within the admissible range, so the polarization shell cannot be said to repair the silicon overgap at all. The two barriers are thereby priced separately: the basis barrier is the larger of the two only before the on-site convention is corrected, and what survives the correction is a residual of one to two eV that a polarization shell closes partly and by a convention-dependent amount, while the $l$-channel dependence admits no remedy within a shared local potential by \cref{prop:nonmult}. Together they explain why beyond-semilocal gap physics has entered tight binding only through explicit Hamiltonian extension rather than through the tables, a connection we develop in the Discussion.

\begin{table}[tb]
\caption{One $d$ shell removes most of the overgap at fixed on-site energies, and what it leaves behind grows with polarity. Gaps are reported at the confined-atom on-site convention, held fixed between the $sp$ and $spd$ sets so that this table isolates the basis-completeness effect alone. Both columns are $\Gamma$-point gaps for parent and model alike, so this is an overgap decomposition and not a transfer ratio. The free-atom closure and its dependence on the $d$-channel placement are given in \cref{fig:dconv}.}\label{tab:spd}%
\setlength{\tabcolsep}{2.5pt}\footnotesize
\begin{tabular*}{\textwidth}{@{\extracolsep{\fill}} l
  S[table-format=1.2]
  S[table-format=2.2] S[table-format=+2.2]
  S[table-format=2.2] S[table-format=+1.2] S[table-format=1.2]
  S[table-format=2.0] @{}}
\toprule
 & {Parent} & \multicolumn{2}{c}{Minimal $sp$ (eV)} & \multicolumn{3}{c}{$d$-polarized $spd$ (eV)} & {Closure} \\
\cmidrule(lr){3-4}\cmidrule(lr){5-7}
System & {DFT (eV)} & {Gap} & {Overgap\tnm{1}}
       & {Gap} & {Residual\tnm{2}} & {$\times$ par.} & {(\%)\tnm{3}} \\
\midrule
Diamond & 5.20 & 14.97 & +9.78  & 6.83  & +1.63 & 1.31 & 83 \\
Silicon & 2.28 & 11.81 & +9.53  & 4.71  & +2.43 & 2.07 & 75 \\
3C-SiC  & 5.54 & 16.89 & +11.34 & 10.46 & +4.92 & 1.89 & 57 \\
\addlinespace
Mean & {--} & {--} & {--} & {--} & {--} & {--} & 71 \\
\bottomrule
\end{tabular*}
\tabnote{\textsuperscript{1}\,The published-set baseline of \cref{fig:bands}d--f, run on these solids in the same DFTB+ build, calibrates the overgap against fitted tables: the excess is shared with fitted parameterizations rather than specific to our convention. \par\smallskip \textsuperscript{2}\,Intrinsic to the two-center on-site approximation, and not removed by enlarging the basis further. It is largest for the polar carbide, and even after the repair the gap remains $1.3$ to $2.1$ times the parent. \par\smallskip \textsuperscript{3}\,Fraction of the $sp$ overgap removed by the added $d$ shell. Confinement radii and on-site eigenvalues are held fixed between the two table sets, so the valence $l$-set is the only change. Overgaps and closures are computed from the unrounded gaps and may differ in the last digit from a subtraction of the rounded values shown.}
\end{table}

Band structures and densities of states along L--$\Gamma$--X--W--K--$\Gamma$ for the three solids, under PBE and CIDER23X ($sp$) and PBE ($spd$) in the free-atom convention, are shown in \cref{fig:bands}a--c; the band-path $\Gamma$-gaps reproduce the free-atom $\Gamma$-gaps above to $\le0.01$~eV. Two table families exist for the carbide and must not be confused: the released parameter sets, which are authoritative throughout and give $6.847$~eV (PBE) and $2.915$~eV (CIDER23X), and the electronic-only tables built for the basis test above, which differ for 3C-SiC alone and are used nowhere else. A lattice-constant scan excludes geometry as the cause of the difference. For diamond the CIDER23X gap is direct at $\Gamma$ in this minimal-$sp$ basis, an artifact of the $sp$-only tables (real diamond is indirect), so its fundamental gap coincides with the $\Gamma$-point value. All gaps quoted here are taken at a common self-consistent charge set.

\subsection{What inherits: the ionic--covalent split}\label{sec:r4}

The transfer failure is not uniform across chemistry, and its structure is a design rule: proceed today for ionic and closed-shell chemistry, where both repulsive potentials and rocksalt-oxide gaps inherit, and extend the channel or the basis first for the elemental and III--V covalent networks, which fail in both observables. The oxide networks sit between the two, their pair potentials transferring as well as any in the study while their gaps do not inherit at all. We resolve the split along those two observables, and they classify by different criteria: a repulsive \emph{pair} is classed by the ionicity of its bond, whereas a \emph{compound} is classed by its network topology, so a pair classed ionic can belong to a compound classed as a covalent network.

\begin{figure}[tb]
\centering
\includegraphics[width=0.90\textwidth]{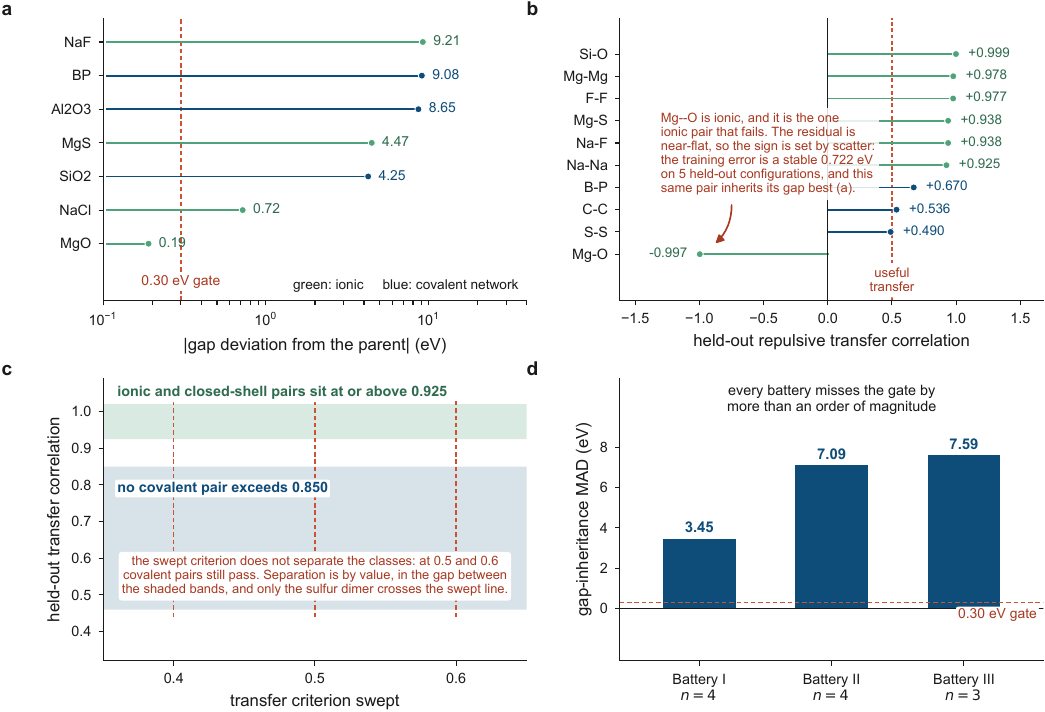}
\caption{The bonding-class structure of the transfer failure.
\textbf{a}, Gap inheritance per compound, against the $0.30$~eV gate. Only the rocksalt oxide and chloride approach it; every covalent network diverges, and two ionic compounds do so as well, by distinct minimal-basis mechanisms.
\textbf{b}, Held-out repulsive-potential transfer, scored as the mean-centered correlation between a potential fitted on one configuration set and tested on another, separating cleanly by bonding class. Mg--O is drawn in its own class colour as the one ionic pair that fails; the text diagnoses it as a near-flat residual rather than a fitting instability.
\textbf{c}, The class separation and the swept criterion, which are different things. The shaded bands are the two class ranges; the gap between them is where a class-separating bar must fall. The swept criterion lies inside the covalent band and does not separate the classes, moving only the sulfur dimer.
\textbf{d}, The per-battery means against the gate.
}
\label{fig:split}
\end{figure}

Held-out repulsive-potential transferability, scored as the correlation of a potential fitted on one configuration set and tested on another, separates by bond class across the sixteen scored pairs (\ref{sn:repulsive}): the two classes are disjoint in value, every covalent pair falling at or below $0.850$ and every ionic and closed-shell pair except Mg--O at or above $0.925$, with the sulfur dimer lowest of all (\cref{fig:split}b). The separation is a property of the chemistry, but it is not resolved by the $0.5$ criterion used elsewhere in this work: at that gate five of the six covalent pairs still pass, and the classes are cleanly separated only by a bar in the interval $(0.850,\,0.925]$. Sweeping the criterion across $0.4$ to $0.6$ moves a single pair, the sulfur dimer (\ref{sn:threshold}). No comparable parameterization study scores repulsive transferability on held-out configurations at all. For the gaps the criterion matters even less, since no solid comes close to the threshold from either side.

The magnesium-oxide pair is an anti-transfer outlier, a correlation of $-0.997$ on five held-out configurations against a stable $0.722$~eV training error, which is the signature of a near-flat repulsive residual rather than a fitting instability. A single distance-dependent repulsive cannot span the coordination changes of a covalent network, whereas a closed-shell pair samples a narrow bonding geometry that a single potential covers.

The second observable, electronic gap inheritance, is dominated by the minimal-basis overgap and inherits only in the most ionic cases (\cref{fig:split}a). Measured as the deviation of the DFTB gap from its parent across eleven compounds in three test batteries, the per-battery mean absolute deviations are $3.45$, $7.09$ and $7.59$~eV, all an order of magnitude above the $0.30$~eV gate (\cref{fig:split}d). Only the battery-one rocksalt oxide and chloride approach the gate; every covalent network diverges, and so do the two remaining ionic compounds by distinct minimal-basis mechanisms, magnesium sulfide as an ionic overgap and sodium fluoride as a confinement-driven undergap.

The two observables therefore agree only at the rocksalt extreme. The closed-shell rocksalt oxide is the one chemistry that inherits in both, and the elemental and III--V covalent networks fail in both. Elsewhere they diverge, and the divergence follows from the two classification criteria being distinct: Si--O and Al--O are the two best repulsive transfers in the study ($0.999$ and $0.998$), yet the oxide networks they build, SiO$_2$ and Al$_2$O$_3$, are among the worst gap inheritors ($4.25$ and $8.65$~eV). A well-transferring pair potential samples a narrow bonding geometry; the gap of the network it builds is set by the minimal-basis conduction manifold, which that potential does not touch. Two of the four ionic compounds also fail the gap gate, magnesium sulfide at $4.47$~eV and sodium fluoride at $9.21$~eV, so the ionic side of the gap split rests on magnesium oxide and sodium chloride alone.

Feasibility of the pipeline itself extends across the periodic table: counting every distinct element that passes across all stages gives a twenty-three-element parameterization map. The binding constraint is the periodic machine-learned self-consistent field, bounded for indium, tin, and antimony.

\subsection{Application boundary: excited states and defect levels}\label{sec:r5}

\begin{figure}[tb]
\centering
\includegraphics[width=0.90\textwidth]{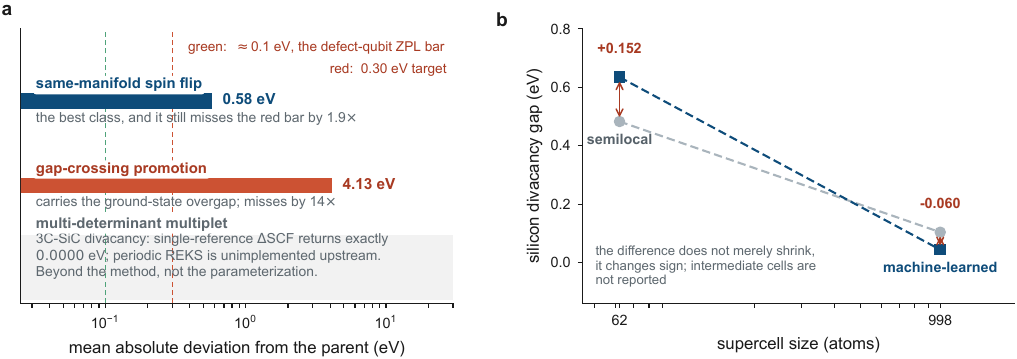}
\caption{The application boundary: what the compressed model can and cannot be asked for.
\textbf{a}, Three regimes on a common scale of deviation from the parent: same-manifold spin flips track it closely, gap-crossing promotions carry the ground-state overgap with them, and multi-determinant multiplets return an identically zero splitting under single-reference $\Delta$SCF, which is a method boundary rather than a parameterization one.
\textbf{b}, The silicon divacancy gap against supercell size, the inter-functional difference changing sign as the cell grows.}
\label{fig:boundary}
\end{figure}

The same guidance extends to excited states, and here it is an ordering, not an absolute licence: $\Delta$SCF fidelity against the parent is structured by transition character, with same-configuration spin flips reproducing the parent to a mean absolute deviation of $0.58$~eV against $4.13$~eV for gap-crossing promotions that carry the ground-state overgap (\cref{fig:boundary}a). The $0.58$~eV spin-flip deviation still clears no external target, sitting well above both the $\approx0.1$~eV defect-qubit zero-phonon-line tolerance and the $0.30$~eV inheritance gate, so the actionable content is the relative ordering, not a usable absolute accuracy. Because the two functionals were run on different diatomic sets (only O$_2$ shared), these deviations are a per-functional fidelity check of DFTB against its own parent rather than a measurement of inheritance.

Put to a real defect, that class meets a further, method-level wall: for the silicon carbide divacancy \citep{koehl2011,zalandauskas2025divacancy} the degenerate-$e$ manifold makes the transition multi-determinant, and single-reference $\Delta$SCF returns a $0.0000$~eV splitting under both functionals, so no zero-phonon line is claimed \citep{filatov2015reks}. The surrounding defect-thermodynamics machinery nonetheless runs on the single-reference silicon divacancy, where the difference between the two functionals' charge-transition levels changes sign as the cell grows, from $+0.152$~eV at $62$ atoms to $-0.060$~eV at $998$ (\cref{fig:boundary}b), so the functional's effect on the defect decouples with size. The level itself stays quantitatively bounded by the same overgap that limits the bulk gaps; the supercell-convergence protocol, the extrapolated level, and the direct-versus-indirect caveat are given in the Supplementary excited-state note (\ref{sn:excited}).

\subsection{An independent meta-GGA control}\label{sec:r2scan}

The anti-transfer is not peculiar to a machine-learned parent or to our inversion bridge. Pushing a conventional meta-GGA~\citep{furness2020r2scan}, r$^2$SCAN~\citep{furness2020r2scanerratum}, through the identical parent~$\to$~confined-atom~$\to$~Slater--Koster~$\to$~DFTB pipeline reproduces the same wrong-sign result (Supplementary \cref{tab:r2scan}). At the parent level r$^2$SCAN opens every gap over PBE, yet the compressed DFTB gaps close, giving a negative free-atom transfer ratio for each of the three solids (diamond $-1.049$, silicon $-2.674$, 3C-SiC $-0.442$), the same sign as CIDER23X, and negative under both on-site conventions. One caveat bounds what this control isolates: our two-center integrator cannot build r$^2$SCAN tables, so the functional enters only through its on-site eigenvalues placed on shared PBE two-center tables, and the experiment probes the on-site-convention mechanism rather than the $f$-channel of the theorem directly. The comparison is nonetheless clean, because the two-center construction is held fixed while the on-site set is the only thing that changes. What the control shows is that a standard meta-GGA fails the gap test the same way a machine-learned functional does, consistent with the obstruction being a property of orbital dependence in general.

\section{Discussion}\label{sec:discussion}
Compression through the standard parameterization channel preserves what lives in the occupied manifold and in orbital geometry, and discards what lives in the fundamental gap. We propose the transfer ratio as a standard pre-test: before a machine-learned functional is adopted as the parent of a tight-binding parameterization campaign, the ratio, evaluated on a small set of test solids, indicates whether the property of interest is likely to be inherited, at negligible cost relative to the campaign itself.

The reason the gap is the casualty is structural, with an implication for the direction of functional development. Orbital dependence lies outside the representational capacity of the standard channel, yet it is precisely the information that modern machine-learned functionals introduce to improve band gaps. If the orbital-dependent channel carries the bulk of a functional's gap correction, a condition \cref{prop:gap} states and this study does not evaluate, then improvement and compressibility are in tension rather than merely uncorrelated: the more a functional relies on higher-rung orbital information to correct gaps, the less faithfully the channel can represent that correction, and to climb Jacob's ladder is to acquire orbital dependence rather than to shed it. The tension is therefore a conditional expectation rather than an established property of the class. Under the physically motivated free-atom on-site convention it is sharpest, the correction anti-transferring rather than simply failing to transfer, so that across every covalent solid considered a more accurate parent yields a less faithful tight-binding representation of its own band gap.

The r$^2$SCAN control shows the obstruction is not specific to machine-learned parents: a conventional meta-GGA anti-transfers in every solid and under both on-site conventions, consistent with orbital dependence itself, rather than machine learning, being the obstruction.

The second barrier is separate, and it is the one that admits a repair within the existing Hamiltonian form, though the repair is partial. Most of what presents as a basis deficiency is on-site bookkeeping; correcting it removes the bulk of the excess and leaves a residual of one to two electronvolts that a polarization shell closes only in part, and by an amount that the placement of the empty $d$ level fixes rather than the physics. Read against that decomposition, the established beyond-semilocal tight-binding schemes, which add explicit long-range or hybrid exchange to the Hamiltonian, and the extended-basis parameterizations that document and repair the minimal-basis conduction-band deficiency, have been meeting both costs empirically and at once. Our contribution is to price them separately and to identify which one forces a change of Hamiltonian form.

The optimized effective potential~\citep{yangwu2002oep,kummelperdew2003oep} is the natural objection. For an orbital-dependent functional it constructs the best single multiplicative potential, and our production inversion is a practical stand-in for it, so the measured anti-transfer already states what the best local representative delivers. What it cannot do is restore the orbital-dependent part of the response, where the orbital-dependent part of the gap correction lives, so the extensions below change the form of the Hamiltonian rather than the potential entering it.

The bonding-class split defines the immediate opportunity. Repulsive-potential transfer is clean for ionic and closed-shell pairs, and gap inheritance approaches its target for the rocksalt oxides, placing functional-consistent tight binding on its firmest footing for closed-shell oxide chemistry. Covalent networks require Hamiltonian or basis extensions before comparable transfer becomes possible, and the requirement is specific to the gap: the oxide networks already inherit their repulsive potentials, so what has to be built is a conduction-manifold remedy rather than a better two-body fit. The obstruction arises from how covalent networks form their gaps, not from a universal limitation of functional-consistent tight binding.

Our results also distinguish two research programs that are often conflated. The data-driven program fits tight-binding tables directly to reference quantities such as band structures or local environments, and a model fitted to more accurate reference bands will generally improve; nothing here contradicts that, and the theorem does not apply, because fitted tables are not constrained to arise from a local potential. Our conclusions instead concern the functional-consistent program, in which the tight-binding tables are derived directly from the parent functional; there, where the objective is a physics-derived model that inherits both the approximations and the predictive behavior of its parent, orbital dependence becomes a fundamental obstacle that fitting bypasses but derivation cannot.

Several boundaries define the scope of these conclusions. The tables are intentionally left unfitted so that the parent functional is the only varied ingredient, which makes the reported overgaps larger than those of production parameterizations and confines the conclusions to the functional-consistent regime rather than to empirically tuned DFTB. The periodic benchmark comprises four covalent solids and one ionic control, and the molecular proxy three diatomics, so every reported average carries its sample size and is not a distributional statement; CIDER23X carries the periodic route and CIDER24Xe the molecular proxy, and both reach the same conclusion. The conclusions are insensitive to the confinement convention.

Because the method already succeeds for closed-shell and ionic systems, future work should extend it to the covalent networks where the present formulation fails, with the transfer ratio as the quantitative acceptance criterion for every proposed extension. We see three natural directions: a uniform polarization and double-$\zeta$ basis, which addresses the overgap directly and for which the extended-basis result of \cref{tab:spd} is already proof of principle; $l$-dependent on-site potentials, which relax the single-local-potential assumption and so address the non-representability result most directly; and explicit environment-dependent two-center terms, the deepest option. Emerging machine-learned functionals with semilocal cost are candidates for the same inversion framework once periodic implementations appear, and we expect the same structural obstruction, a prediction the r$^2$SCAN control already supports. 

\section{Methods}\label{sec:methods}

\subsection{Atomic references and confinement}
We solved confined pseudo-atoms with a PySCF-based radial Kohn--Sham solver \citep{pyscf2020} on a level-6 radial grid, using even-tempered valence bases of 16 functions for hydrogen, 24 for carbon and nitrogen, and 28 for oxygen and silicon, with a $3d$ polarization channel on silicon. Confinement was a power-2 potential of radius $r_0 = 1.85\,r_{\mathrm{cov}}$ from the Cordero covalent radii \citep{cordero2008}, with hydrogen fixed at $3.0$~Bohr and silicon at $3.30$~Bohr by the overlap criterion of the Results. Elements heavier than sulfur were treated scalar-relativistically at the spin-free exact-two-component (sfx2c1e) level \citep{liupeng2009}. From the same solver we obtained on-site Hubbard $U$ values as finite-difference second derivatives of the free-atom energy with respect to occupation, required to vary smoothly with occupation for each functional, together with shell-resolved spin constants.

\subsection{Potential inversion and table generation}
For each functional we assembled the confined-atom effective potential directly where a closed-form semilocal expression exists, and otherwise recovered it by radial Kohn--Sham inversion of the confined orbitals~\citep{wuyang2003} with node preservation and spherical-harmonic projection for signed-orbital recovery. Channel-asymmetry percentiles are reported under a node mask retaining only radii at which both $|u_s|$ and $|u_p|$ exceed $5\%$ of their maxima, outside which the inversion is ill-conditioned. The recovered potential was injected into hotcent~\citep{hotcent} through its two-center offsite table interface, producing two disjoint table sets. Overlap-regression tables use density superposition, matching the mio convention, and serve only to certify the overlap $S$ against the reference mio parameterization. Production electronic tables, which underlie every gap, transfer-ratio, and overgap result, use the imported-potential (potential-superposition) Hamiltonian, the route through which an inverted machine-learned potential enters the tight-binding problem.

\subsection{Parent DFT}
Molecular parent calculations used PySCF with the def2-TZVP basis \citep{weigend2005def2}. Periodic parent calculations used GPAW \citep{gpaw2010} with the projector augmented-wave method, a $500$~eV plane-wave cutoff, and an $8\times8\times8$ $\Gamma$-centered $k$-mesh for diamond, silicon, and 3C-SiC. These settings are converged: the parent gap changes by $\le0.003$~eV from $500$ to $700$~eV cutoff and by $\le0.0006$~eV from $8^3$ to $12^3$ $k$-points, the DFTB integration grid and self-consistent-charge tolerance ($10^{-4}$ to $10^{-6}$) change the $\Gamma$-gap by $\le0.001$~eV, and the production values reproduce the frozen gaps to $0.0006$~eV (diamond) and $0.0015$~eV (silicon). Structures were built with the Atomic Simulation Environment \citep{ase2017} at Materials Project reference lattice constants \citep{materialsproject2013}. The semilocal reference functional was PBE \citep{perdew1996pbe}; machine-learned functionals were evaluated through CiderPress \citep{ciderpress} with the published functional definitions and per-set $k$-mesh, smearing, and mixer settings.

\subsection{DFTB}
DFTB calculations used DFTB+ \citep{dftbplus2020}: a conda-forge serial build for molecular tests, a from-source build linking ELSI \citep{elsi2018,elsi2020} with ELPA \citep{elpa2014} and NTPoly \citep{ntpoly2018} for large cells, and an ARPACK-enabled rebuild \citep{arpackng} for the $\Delta$SCF work. Dense parallel diagonalization (ELPA) was assigned to defect quantities and the linear-scaling solver (NTPoly) to gapped bulk. Self-consistent charges were converged to $10^{-5}$~$e$ on the Mulliken charge. Defect supercells used $\Gamma$-only sampling, and Fermi smearing at $300$~K ($k_{\mathrm{B}}T \approx 0.026$~eV) was applied to the metallic-prone defect cells and switched off for the gapped bulk.

\subsection{Transfer-ratio protocol}
The transfer ratio is the DFTB property change divided by the parent-DFT property change under a change of functional only. We evaluated it for the fundamental gap and the occupied bandwidth on diamond, silicon, 3C-SiC, and boron phosphide, with MgO as ionic control and N$_2$, CO, and CO$_2$ as molecular proxies, at lattice constants held fixed at the Materials Project values ($a = 3.567$~\AA{} for diamond, $5.430$~\AA{} for silicon, and $4.360$~\AA{} for 3C-SiC) across both functionals so that only the functional varies. One gap definition is used on both sides of every transfer ratio: the fundamental gap, taken as the minimum over the sampled band path, at the parent-DFT level and at the DFTB level alike. The distinction matters because for silicon the $\Gamma$-point gap exceeds the fundamental gap by more than an electronvolt, so a ratio formed from a fundamental-gap denominator and a $\Gamma$-point numerator would divide one observable by another. DFTB charges were converged on a $12^3$ Monkhorst--Pack mesh and the eigenvalues then evaluated along the $L$--$\Gamma$--$X$--$W$--$K$--$\Gamma$ path at fixed charges; a decision threshold of $0.5$ was applied, and compounds that failed to converge were dropped rather than adjusted. The overgap decomposition of \cref{tab:spd} is a separate quantity and is not a transfer ratio; it is reported at $\Gamma$ for both parent and model. The only systems dropped were the periodic machine-learned self-consistent fields of indium, tin, and antimony, three of the twenty-three screened elements, which failed the periodic PAW-setup control-iteration gate (non-convergent at $10$, $16$, and $13$ control iterations); because their atomic-feasibility gates pass, they enter the coverage map but not the periodic transfer battery. No molecular or bulk transfer-ratio system was dropped.

\subsection{Extended-basis test}
We rebuilt the carbon and silicon electronic tables with an added $d$-polarization channel ($spd$), holding the confinement radii and on-site eigenvalues identical to the minimal-$sp$ tables so that only the valence $l$-set changes, and re-measured the $\Gamma$-point gaps of diamond, silicon, and 3C-SiC with both table sets against parent gaps computed with identical settings. The minimal-$sp$ pipeline reproduced the frozen baseline (silicon $11.811$ against $11.808$~eV; diamond $14.973$~eV exactly). The periodic gate of the main comparison is strictly minimal-$sp$: the silicon $3d$ channel enters only the O--Si overlap certification, never the periodic Hamiltonian (\ref{sn:spd}).

\subsection{Repulsive fitting}
Two-body repulsive potentials were fitted with curvature-constrained splines (CCS 0.22.5) \citep{ccs2021} to reference configurations combining rattled cells, equation-of-state scans, and dimer bond scans under a train/test split. Transferability was scored on held-out configurations with a mean-centered root-mean-square error and a Pearson correlation, removing offset dependence.

\subsection{Defects and \texorpdfstring{$\Delta$}{Delta}SCF}
The solver-scaling series used $\Gamma$-only defect supercells with Fermi filling at $300$~K and no ionic relaxation; the transition-level series used charged, relaxed cells with a monopole (Makov--Payne) image correction \citep{makov1995}. A potential-alignment term was tested and rejected because it did not improve the $1/L\to0$ extrapolation. Excited-state $\Delta$SCF \citep{kowalczyk2011} used non-Aufbau occupations with spin purification and the shell-resolved spin constants above. The $\Delta$SCF test sets differ between functionals (N$_2$, CO, and O$_2$ for the semilocal reference; O$_2$, SiO, and Si$_2$ for the machine-learned functional; only O$_2$ in common), so these deviations measure the fidelity of DFTB against its own parent per functional rather than what a change of functional inherits; we therefore report only the ordering by transition character and leave a common-set rerun, which would license an inheritance reading, to future work.

\subsection{Error control}
Every calculation reported here is deterministic, with no training, sampling, or random seed, so all uncertainty is numerical and systematic rather than statistical. We bound it with four independent tests. Basis and grid convergence is established at the atomic level by the sub-microhartree plateau of the free-atom energy. Sensitivity to the one free convention, the confinement radius, is bounded by sweeping the carbon reference from $2.0$ to $7.0$~Bohr, which leaves the overgap and the $l$-channel asymmetry unchanged. Sensitivity to the one free decision threshold is bounded by re-scoring at $0.4$, $0.5$, and $0.6$; the gap conclusion is threshold-independent outright, since no solid reaches a transfer-ratio magnitude of $0.4$ under the confined convention and every covalent ratio is negative under the free-atom convention. Solver and finite-size error are bounded by the agreement of three eigensolvers to full precision and by supercell convergence to $998$ atoms. Sample sizes are stated wherever a mean is quoted (four covalent solids, one ionic solid, and three molecules for the transfer ratio; eleven compounds for gap inheritance; sixteen pairs for repulsive transfer; six diatomics for $\Delta$SCF), and no mean is offered as a distributional estimate.


\begin{compute}
Local analysis ran on an Apple M4 Pro (24~GB); periodic and machine-learned reference calculations ran on a single RTX 4090 GPU node. The full project ledger totals $290$ GPU jobs and $92.5$ active GPU-hours; the inversion bridge itself is a one-time per-element atomic solve of seconds to minutes, amortized across every downstream table, so the pipeline cost is dominated by the parent-DFT references, as in any conventional parameterization (\ref{sn:compute}).
\end{compute}

\begin{contributions}
C.P. conceived and implemented the parameterization bridge and functional-agnostic pipeline, conducted all computational simulations and benchmark experiments, performed the data analysis, and wrote the original manuscript. E.S., M.K., and H.K. supervised the research, contributed to the interpretation of results, and reviewed and edited the manuscript. All authors read and approved the final manuscript.
\end{contributions}

\begin{funding}
No funding was received for this research.
\end{funding}

\begin{availability}
The frozen data underlying every figure and table, the certified Slater--Koster parameter sets, and the release manifest with checksums are openly available at \url{https://github.com/KurbanIntelligenceLab/mlxc-dftb}, where the twenty-five parameter files are published under \texttt{skf/} carrying free-atom on-site energies in their homonuclear blocks, with a manifest recording the checksum and on-site convention of every file. The contents of each released set are itemized in \ref{sn:release} (Supplementary \cref{tab:release}), and a single-command reproducibility suite regenerates every reported result.

The \texttt{mlxc\_dftb} package and the validation workflow are openly available at \url{https://github.com/KurbanIntelligenceLab/mlxc-dftb}.
\end{availability}

\begin{conflicts}
The authors declare no competing financial or non-financial interests.
\end{conflicts}

\bibliography{references}

\clearpage
\setcounter{section}{0}
\setcounter{table}{0}
\setcounter{figure}{0}
\renewcommand{\theHsection}{SI.\arabic{section}}
\renewcommand{\theHtable}{SI.\arabic{table}}
\renewcommand{\theHfigure}{SI.\arabic{figure}}
\renewcommand{\theHequation}{SI.\arabic{equation}}
\renewcommand{\thesection}{Supplementary Note \arabic{section}}
\renewcommand{\thetable}{S\arabic{table}}
\renewcommand{\thefigure}{S\arabic{figure}}

\begin{center}
{\kilsemi\large\color{kilink}Supplementary Information}\\[10pt]
{\kilsemi\LARGE\color{kilink}\par\setlength{\baselineskip}{1.25\baselineskip}When do machine-learned exchange--correlation improvements inherit into density-functional tight binding?\par}
\end{center}
\bigskip

\section{Derivations}\label{app:theory}

This Supplementary Note carries the complete derivations of the theoretical results stated in the main text. The statements themselves are in the main text; what follows is their support. Everything here is standard analysis applied to the generalized Kohn--Sham equation of an orbital-dependent functional; we give it in full because the parameterization channel's error is an equality, not a bound, and the equality is what makes those predictions sharp.

\subsection{Notation}\label{app:notation}

\begin{center}
\begin{tabular}{@{}ll@{}}
\toprule
Symbol & Meaning \\
\midrule
$\tau$ & kinetic energy density, $\tfrac{1}{2}\sum_i^{\mathrm{occ}}|\nabla\psi_i|^{2}$ \\
$e_{\mathrm{xc}}(n,\nabla n,\tau)$ & exchange--correlation energy density \\
$f = \partial e_{\mathrm{xc}}/\partial\tau$ & dimensionless; zero at the semilocal rungs \\
$v(\mathbf{r})$ & multiplicative part of the effective potential \\
$S_{nl}(r) = r^{-l}R_{nl}(r)$ & reduced radial function, $S_{nl}(0)\neq 0$ \\
$\phi_\mu$ & confined atomic orbital, a two-center basis function \\
$\chi_v,\ \chi_c$ & valence- and conduction-edge eigenstates of the \emph{solid} \\
$Z$ & nuclear charge \\
$H^{\mathrm{ch}}, H^{\mathrm{GKS}}$ & channel and exact Hamiltonian matrix elements \\
$\Delta H = H^{\mathrm{GKS}} - H^{\mathrm{ch}}$ & the residual of the compression \\
\bottomrule
\end{tabular}
\end{center}

\subsection{Assumptions}\label{app:assump}

Every theoretical result of the main text rests on the following, and we state them here rather than inside a proof.

\begin{description}
\item[A1.] The confined atom is spherical and spin-unpolarized, so that $v$ and $f$ are functions of $r$ alone and the orbitals separate as $R_{nl}(r)Y_{lm}$.
\item[A2.] $e_{\mathrm{xc}}$ is differentiable in $\tau$ and $f = \partial e_{\mathrm{xc}}/\partial\tau$ is continuous on the radial domain considered.
\item[A3.] The $R_{nl}$ are exact solutions of the radial generalized Kohn--Sham equation, and the $l$-resolved local representative is evaluated away from the nodes of $R_{nl}$.
\item[A4.] $f\,\phi_\mu\,\nabla\phi_\nu \to 0$ on the boundary of the integration domain, so that the integration by parts below carries no surface term. This holds for confined orbitals, which vanish outside the confinement radius.
\end{description}

\noindent Assumption A3 carries most of the weight and is an idealization: the $s$ channel has a radial node and the $p$ channel does not, so the numerical error in $v_s$ is not the numerical error in $v_p$ and a measured $|v_s - v_p|$ mixes the physical $l$-dependence with an artifact of the same qualitative signature. That is why prediction P1 is required and not optional: under PBE the physical term vanishes identically, so any residual $|v_s - v_p|$ is the artifact, measured directly. A2 fails at a nucleus for functionals whose $f$ is discontinuous there; none of those considered here is.

\subsection{The generalized Kohn--Sham equation}

Varying $E_{\mathrm{xc}}[n,\nabla n,\tau]$ with respect to $\psi_i^{*}$, using $\delta\tau/\delta\psi_i^{*} = -\tfrac{1}{2}\nabla^{2}\psi_i$ in the weak sense, and collecting the multiplicative contributions into $v$ gives the generalized Kohn--Sham equation of the main text,
\begin{equation}\tag{\ref{eq:gks}}
-\tfrac{1}{2}\,\nabla\!\cdot\!\big[(1+f)\,\nabla\psi_i\big] + v\,\psi_i = \varepsilon_i\,\psi_i ,
\end{equation}
the standard form for an orbital-dependent functional \citep{perdew1999metagga,kummel2008oep,bystrom2024nonlocal}, equivalently a position-dependent effective-mass problem with $m^{*}(r) = [1+f(r)]^{-1}$. Nothing in it is ours.

\subsection{Proof of \texorpdfstring{\cref{prop:vl}}{the l-resolved local representative}}\label{app:vl}

Under A1, write $\psi = R_{nl}(r)Y_{lm}$ and $g = 1+f$. For a radial $g$,
\begin{equation}
\nabla\!\cdot\!\big(g\,\nabla\psi\big) = \left[\frac{1}{r^{2}}\big(r^{2} g\, R'\big)' - g\,\frac{l(l+1)}{r^{2}}\,R\right] Y_{lm},
\end{equation}
so the generalized Kohn--Sham equation reduces to
\begin{equation}\label{eq:radgks}
-\frac{1}{2r^{2}}\big(r^{2} g\, R_{nl}'\big)' \;+\; g\,\frac{l(l+1)}{2r^{2}}\,R_{nl} \;+\; v\,R_{nl} \;=\; \varepsilon_{nl}\,R_{nl}.
\end{equation}
The ordinary radial Kohn--Sham equation that the parameterization channel can express, by \cref{def:channel}, is \cref{eq:radgks} with $g \equiv 1$ and $v$ replaced by some multiplicative $v_l$:
\begin{equation}\label{eq:radks}
-\frac{1}{2r^{2}}\big(r^{2} R_{nl}'\big)' \;+\; \frac{l(l+1)}{2r^{2}}\,R_{nl} \;+\; v_l\,R_{nl} \;=\; \varepsilon_{nl}\,R_{nl}.
\end{equation}
Subtracting \cref{eq:radks} from the equation above, using $g - 1 = f$, and dividing by $R_{nl}$, which is permitted away from its nodes by A3,
\begin{equation}
-\frac{1}{2r^{2}R_{nl}}\big(r^{2} f\, R_{nl}'\big)' \;+\; f\,\frac{l(l+1)}{2r^{2}} \;+\; v \;-\; v_l \;=\; 0,
\end{equation}
which rearranges to the $l$-resolved local representative. Uniqueness is immediate: the ordinary radial equation determines $v_l$ pointwise once $R_{nl}$ and $\varepsilon_{nl}$ are fixed. $\square$

\subsection{Affine structure of the representative}\label{app:dvdl}

\begin{proposition}[Affine structure of the representative]\label{prop:dvdl}
Write $R_{nl}(r) = r^{l}\,S_{nl}(r)$ with $S_{nl}(0)\neq 0$. Then $v_l - v$ is exactly affine in $l$ at fixed $S_{nl}$,
\begin{equation}\label{eq:dvdl}
v_l(r) - v(r) \;=\; -\,\frac{l}{2r}\left[\,f'(r) + \frac{2 f(r)\,S_{nl}'(r)}{S_{nl}(r)}\right] \;+\; C[S_{nl}](r),
\end{equation}
where $C$ depends on $S_{nl}$ but not on $l$. The $l(l+1)$ terms of the local representative cancel identically, so it acquires no quadratic $l$-weighting, and every surviving $l$-dependent term is proportional to $f$ or to $f'$.
\end{proposition}

\noindent Substituting the cusp-regular form into \cref{eq:vl} and differentiating with respect to $l$ at fixed $S_{nl}$ gives
\begin{equation}
\frac{\partial v_l}{\partial l} \;=\; -\,\frac{f'(r)}{2r} \;-\; \frac{f\,S_{nl}'}{r\,S_{nl}} ,
\end{equation}
which is independent of $l$, so \cref{eq:dvdl} follows with $C[S_{nl}]$ collecting the $l$-independent part: the term $f\,l(l+1)/2r^{2}$ is cancelled exactly by the $l(l+1)$ piece that $R_{nl}' = r^{l-1}(l S_{nl} + r S_{nl}')$ generates inside the second term. The naive reading, that the representative acquires an $l(l+1)$-weighted piece of $f$, would predict a factor of three between $v_d-v_s$ and $v_p-v_s$ where the affine structure predicts a factor of two. $\square$

\subsection{Proof of the non-representability theorem}\label{app:thm}

If $f\equiv 0$, the local representative gives $v_l = v$ for every $l$, so $v_{\mathrm{ch}} = v$ satisfies \cref{def:rep}. This covers the local and generalized-gradient rungs, on which the parameterization channel has rested for thirty years.

Let $f\not\equiv 0$. By \cref{prop:dvdl} the $l$-dependent part of $v_l - v$ is
\begin{equation}\label{eq:slope}
-\,\frac{l}{2r}\left[f'(r) + \frac{2 f(r) S_{nl}'(r)}{S_{nl}(r)}\right].
\end{equation}
This expression cannot be compared across channels without knowing how $S_{nl}$ varies with $l$, and in general it does. At the nucleus, however, the variation is fixed by the cusp condition. Expanding the radial solution of \cref{eq:radgks} about a nucleus of charge $Z$ gives $R_{nl}(r) = r^{l}\big(1 - Z r/(l+1) + O(r^{2})\big)$ up to normalization, hence
\begin{equation}
\frac{S_{nl}'(0)}{S_{nl}(0)} \;=\; -\,\frac{Z}{l+1}.
\end{equation}
This condition is the leading order of the exact balance: retaining the effective-mass factor $1+f$ in the $r^{l-1}$ expansion gives $S_{nl}'(0)/S_{nl}(0) = -[Z + (l/2)f'(0)]/[(l+1)(1+f(0))]$, which reduces to the standard condition precisely when $f(0) = f'(0) = 0$.\footnote{The corrections shift the coefficients below at the percent level for the functionals considered here, by $0.46\%$ at $l=1$ and $0.72\%$ at $l=2$ with the measured carbon values ($f(0)=0.0117$, $f'(0)/2Z=0.035$). The corrected $l=1$ and $l=2$ coefficients vanish simultaneously only at $f(0)=f'(0)=0$, the semilocal case, so they admit no common zero and the argument below is unaffected.} Substituting the leading-order condition into \cref{eq:slope} and retaining the leading term as $r\to 0$ gives the near-nucleus law of the main text, which vanishes for $l=0$ and does not vanish for $l\geq 1$ unless its bracket does. Taking the difference of the $l=1$ and $l=0$ cases yields \cref{eq:vpvs}. Suppose a channel representative $v_{\mathrm{ch}}$ existed. Then $v_p = v_s = v_{\mathrm{ch}}$ and the left side of \cref{eq:vpvs} is identically zero, forcing $f'(0) = Z f(0)$. That is a single scalar identity between the functional's $\tau$-derivative and the nuclear charge at which it is evaluated. It is not satisfied by any functional in use, and it cannot be satisfied for every element by one functional, since $Z$ varies and $f(0), f'(0)$ are determined by the density the functional itself produces. Excluding it, $v_p \neq v_s$, no $v_{\mathrm{ch}}$ exists, and the functional is not channel-representable. $\square$

\subsection{The coefficients, and why three channels close the escape}\label{app:kappa}

Write $\kappa \equiv f'(0)/Z f(0)$, a dimensionless number determined by the functional on its own converged density at the nucleus of charge $Z$, and let $c_l$ be the coefficient of $1/r$ in the near-nucleus law. Then
\begin{equation}\label{eq:coeffs}
c_s = 0, \qquad
c_p = \tfrac{1}{2}\,(1-\kappa)\,Z f(0), \qquad
c_d = \tfrac{1}{3}\,(2-3\kappa)\,Z f(0),
\end{equation}
each linear in $\kappa$, and three consequences of \cref{eq:coeffs} bound what the theorem does. The vanishing of $c_s$ is unconditional: the $s$ channel carries no $1/r$ divergence for any functional, the one parameter-free statement of the set. The degenerate case of the theorem, $\kappa = 1$, is exactly the zero of $c_p$, so it is a real escape for a confined atom carrying only $s$ and $p$ channels; but $c_d$ vanishes at $\kappa = 2/3$, so no $\kappa$ annihilates both, and an atom carrying $s$, $p$ and $d$ admits no exact local representative for any functional with $f(0)\neq 0$. The ratio $c_d/c_p = \tfrac{2}{3}(2-3\kappa)/(1-\kappa)$ equals $4/3$ at $\kappa = 0$ and at no other value, so the $l/(l+1)$ weights are drawn at $\kappa = 0$ and the $4/3$ is not a parameter-free prediction; the naive $l(l+1)$ reading would force the ratio to $3$ for every $\kappa$ and is excluded outright by the cancellation of \cref{prop:dvdl}. $\square$

\medskip
\noindent The proof draws its strength from the nucleus: two channels whose local representatives differ by a \emph{diverging} term cannot be reconciled by any bounded multiplicative potential, so the obstruction is not a matter of degree. The price is Assumption~A3, since the cusp condition holds for exact solutions of the radial problem and fails for orbitals expanded in a Gaussian basis, which have no cusp, so the theorem is a statement about the functional and a numerical inversion in a Gaussian basis does not automatically inherit it. This, alongside the node ill-conditioning discussed below, is why the semilocal control of prediction P1 is not optional.

\subsection{Proof of the channel-residual identity}\label{app:dH}

By \cref{def:channel} the channel's matrix element is $H^{\mathrm{ch}}_{\mu\nu} = \langle\phi_\mu| -\tfrac{1}{2}\nabla^{2} + v_{\mathrm{sup}} |\phi_\nu\rangle$, while the exact element under the generalized Kohn--Sham equation is $H^{\mathrm{GKS}}_{\mu\nu} = \langle\phi_\mu| -\tfrac{1}{2}\nabla\!\cdot\!\big[(1+f)\nabla\,\cdot\,\big] + v_{\mathrm{sup}} |\phi_\nu\rangle$. The multiplicative parts cancel in the difference, leaving
\begin{equation}
\Delta H_{\mu\nu} \;=\; -\frac{1}{2}\int \phi_\mu \,\nabla\!\cdot\!\big(f\,\nabla\phi_\nu\big)\,\mathrm{d}^{3}r .
\end{equation}
Integrating by parts and discarding the surface term by A4 gives \cref{eq:dH} of the main text. Suppose $\Delta H_{\mu\nu} = \langle\phi_\mu|w|\phi_\nu\rangle$ for some multiplicative $w$ and for every pair $(\mu,\nu)$ in the basis. Taking $\mu = \nu$ and varying $\phi_\mu$ over orbitals with the same density but different gradients, the left side changes and the right side does not, which is a contradiction unless $f$ is constant, in which case $\Delta H = \tfrac{1}{2} f \langle\nabla\phi_\mu|\nabla\phi_\nu\rangle = f\,T_{\mu\nu}$, a rescaling of the kinetic matrix rather than a potential. The residual is therefore not absorbable into any multiplicative potential, and in particular not into a re-tuned confinement radius, a different superposition convention, or a more accurate inversion. $\square$

\subsection{Derivation of the gap-shift expression and a magnitude estimate}\label{app:gap}

Let $\chi_k$ be a normalized eigenstate of the channel Hamiltonian, in the solid-state notation of \ref{app:notation}. Rayleigh--Schr\"odinger perturbation theory in $\Delta H$ gives, to first order,
\begin{equation}
\delta\varepsilon_k \;=\; \langle \chi_k | \Delta H | \chi_k\rangle \;=\; \frac{1}{2}\int f\,|\nabla\chi_k|^{2}\,\mathrm{d}^{3}r \;=\; \int f(\mathbf{r})\,\tau_k(\mathbf{r})\,\mathrm{d}^{3}r,
\end{equation}
with $\tau_k = \tfrac{1}{2}|\nabla\chi_k|^{2}$ the state-resolved kinetic energy density. Applying this at the valence and conduction edges and subtracting gives \cref{eq:dEg}. $\square$

The sign argument is physical and we do not present it as a proof: the conduction-edge state of a covalent solid is antibonding and changes sign between the atoms, so it carries more curvature in the bonding region than the bonding valence-edge state, whence $\tau_c > \tau_v$ there. The inequality is local to the bonding region, the integral runs over the whole cell, and $f$ is not of one sign everywhere. For an ionic solid, whose band edges are not a bonding and antibonding pair of the same orbitals, the argument does not run at all, which is consistent with the ionic compounds being the ones that inherit their gaps.

Three caveats belong next to this statement rather than at the end of the paper. The expansion is first order in $f$, and $f$ is not uniformly small, so the gap-shift expression fixes the sign and the location of the missing physics but not its exact size. The self-consistent-charge cycle acts on the density and can restore part of the shift indirectly, so it is an upper bound on what is lost only in the non-self-consistent limit. And the confined orbitals $\phi_\mu$ themselves change when the functional changes, a second-order effect in the present counting but a real one; the convention question this raises is settled by the on-site test of the main text.

The channel-residual identity also fixes the scale the measured integrals must be compared against: $f$ is dimensionless and of order $10^{-1}$ across the valence region, the gradient overlap $\int \nabla\phi_\mu\cdot\nabla\phi_\nu$ at a bond length is a kinetic-scale matrix element of order $10^{-1}$ to $1$~Ha, so the expected $sp\sigma$ residual is roughly $10^{-2}$ to $10^{-1}$~Ha. The fixed-radius $sp\sigma$ figure of $-6.21$~Ha against a reference $-0.13$~Ha exceeds that estimate by one to two orders of magnitude, and the resolution is that the figure is not an integral of the form the identity bounds. Of the two candidate readings, an anomalously large core $f$ is excluded on physical grounds: for a meta-generalized-gradient exchange functional $f \sim e_{\mathrm{x}}^{\mathrm{LDA}}/\tau^{\mathrm{unif}} \propto n^{-1/3}$, so $f$ is \emph{smaller} in the high-density core, not larger. Node contamination is the operative reading, and it is measured rather than assumed, in the channel-asymmetry note below. The estimate is therefore consistent with the node-masked statistics, which are the ones the conclusions rest on.

Neither the Slater--Koster tables nor any reported result inherits this artefact. The node-region extremum arises in the diagnostic channel-by-channel inversion, which is run to test the theorem and is not the production route: the tables are integrated by hotcent from the single imported potential, on the two-center radial grid, which does not sample a single contaminated radius in place of an integral. Every gap, transfer ratio, and overgap reported here is computed from those tables. This has been verified directly against the released tables. On the $0.02$~Bohr two-center grid of \texttt{CIDER23X\_Si-Si.skf}, spanning $524$ points to $10.48$~Bohr, the tabulated $sp\sigma$ Hamiltonian element ranges over $[-0.338,\,+0.005]$~Ha and takes the value $-0.118$~Ha at the silicon bond length; the semilocal control gives $-0.127$~Ha at the same separation. No entry anywhere in the table approaches $-6.21$~Ha, and the largest magnitude on the whole grid is smaller than that figure by a factor of eighteen. The corresponding carbon tables give $-0.233$~Ha (machine-learned) against $-0.249$~Ha (semilocal) at the carbon bond length. The node-region extremum is therefore confined to the diagnostic inversion and does not enter the released parameterization or any result computed from it.

\subsection{Scope of the theorem}

Coverage extends beyond the $\tau$ channel\label{app:hybrid}: the argument above is written for $\tau$, but it uses only orbital dependence. For a local hybrid such as DM21 \citep{dm21}, whose ingredients include exact-exchange energy densities $e^{\mathrm{HF}}(\mathbf{r})$, the functional derivative contributes a term of the form $-\int \big(\partial e_{\mathrm{xc}}/\partial e^{\mathrm{HF}}\big)(\mathbf{r})\,\hat{K}(\mathbf{r},\mathbf{r}')\,\psi_i(\mathbf{r}')\,\mathrm{d}^{3}r'$, with $\hat{K}$ the exchange kernel, which is manifestly not multiplicative. For the density-matrix features of CIDER24Xe, the gap-trained functional of our molecular proxy \citep{bystrom2024bandgaps}, the derivative acts on the density matrix and likewise yields an integral operator. In each case \cref{def:rep} fails for the same reason and the channel-residual identity carries over with the corresponding operator in place of $-\tfrac{1}{2}\nabla\!\cdot\!(f\nabla\,\cdot\,)$; only the closed form of the residual changes. The $l$-dependence law is specific to the $\tau$ channel, so prediction P2 applies to meta-generalized-gradient parents and not to local hybrids.

Three qualifications bound what the theorem asserts, and the Discussion states each in full: it constrains the \emph{derivation} of tables from a parent functional and not a model \emph{fitted} to a better functional's bands, to which channel representability does not apply; the residual $\Delta H$ remains even for the best local representative, so the measured anti-transfer is a best-case statement; and the minimal-basis overgap is an independent, partly repairable barrier that follows from none of the above.

\section{Overlap regression, storage conventions, and the confinement scan}\label{sn:overlap}

Across all fifteen element pairs of the C/N/O/Si/H base set, the maximum overlap deviation from the reference tables is $\Delta S_{\max}\le 0.026$, largest for the nitrogen homonuclear pair ($0.0261$), with heteronuclear pairs all $\le 0.017$ (largest H--N, $0.0165$).

The confinement radius is fixed by the overlap criterion $r_0 = 1.85\,r_{\mathrm{cov}}$, with two element-specific exceptions established by the same criterion: hydrogen requires $r_0 = 3.0$~Bohr and silicon $r_0 = 3.30$~Bohr, the silicon value lying below the density-compression window because the overlaps pin the tighter wavefunction extent rather than the density radius. Silicon further requires an explicit $3d$ polarization channel ([Ne]\,$3s^2\,3p^2\,3d^0$) to reproduce the O--Si integrals. Two storage conventions are load-bearing: the X--H integral is stored in the $sp\sigma$ slot of the hydrogen-leading ordering, and the $sp\sigma$ integral requires the standard Slater--Koster sign flip, without which $\Delta S_{\max}$ rises from $\sim 0.003$ to $\sim 0.95$. The Hamiltonian used for the overlap regression is built by density superposition, matching the reference mio set; the imported-potential Hamiltonian, which is the one carried into every reported electronic result, uses potential superposition.

The certification is not balanced on a knife-edge. An eleven-point confinement sweep for the carbon reference over $r_0 = 2.0$--$7.0$~Bohr, from $\Delta S_{\max} = 0.0854$ at the lower endpoint to $0.1968$ at the upper, shows a single smooth optimum of the overlap error at the certified radius $r_0 = 2.657$~Bohr ($\Delta S_{\max} = 0.0043$). The certification is on the overlap; the Hamiltonian root-mean-square difference from the reference tables at that radius is $0.1696$~Ha and is minimized at a different radius, so the tables reproduce the reference $S$ and not the reference $H$, which is consistent with their unfitted construction and is the reason the overgap is a property of this convention rather than of DFTB. The two criteria do not coincide: the overlap error is minimized at the certified radius while the Hamiltonian residual is minimized at $r_0 = 5.000$~Bohr, a factor of $1.9$ away ($H_{\mathrm{rms}} = 0.0806$~Ha there against $0.1696$~Ha at the certified radius). The confinement radius is fixed by the overlap criterion alone, because $S$ is the geometry-independent, convention-free quantity that anchors the basis to the reference mio/pbc parameterization; the tight-binding Hamiltonian is then determined by the imported inverted potential rather than tuned to its own residual minimum, so no free knob is fitted to the electronic result reported here. Certifying against $S$ and letting $H$ follow is the standard functional-consistent construction, and it is why the reported anti-transfer cannot be an artifact of Hamiltonian fitting. The optimum is broad, with a window of roughly $\pm 10\%$ ($\approx 2.4$--$2.9$~Bohr) keeping $\Delta S_{\max}$ below $\sim 0.03$, and smooth degradation on either side. No confinement choice anywhere in the range alters the qualitative conclusions of the main text: the minimal-basis overgap and the $l$-channel nonlocality wall are independent of $r_0$, so the anti-transfer is intrinsic to the standard channel rather than a byproduct of the convention.

\section{The bridge derivation and the ruled-out routes}\label{sn:bridge}

This note reports the empirical route selection and the measured $l$-resolved potentials, which are what predictions P1 and P2 of the main text are tested against. The derivations behind the definitions, the theorem, the propositions and the corollary are given in \ref{app:theory} and not repeated here.

The bridge converts a converged confined-atom Kohn--Sham solution into an effective local potential that the two-center integrator consumes. That integrator refuses meta-GGA functionals outright: its libxc wrapper evaluates the XC energy and potential from the density and its gradient alone, without the kinetic-energy density, so it cannot build r$^2$SCAN two-center tables, and in the r$^2$SCAN control the functional enters only through its on-site eigenvalues, computed with a meta-GGA-capable atomic solver and placed on PBE two-center tables shared identically by both sets. This is a limitation of the integrator we use, not of Slater--Koster parameterization in general, since SkProgs exposes meta-GGA functionals through libxc \citep{skprogs}. Two validations bound the fidelity. The Kohn--Sham inversion the bridge relies on (route B) recovers hotcent's stored effective potential to a median absolute difference of $0.000235$~Ha ($6.4$~meV) over $r\in[0.3,8]$~Bohr. A separate semilocal reconstruction of the exchange--correlation potential from the density (route A) reproduces the PBE truth to $0.000347$~Ha ($9.5$~meV), while the density-gradient-only approximation to that reconstruction fails at $0.0275$~Ha ($748$~meV), two orders of magnitude worse; the Fock-projection route was ruled out for the same reason. Kohn--Sham inversion is the route carried forward. For silicon, inverting the $3s$ and $3p$ confined solutions of the machine-learned functional separately yields local potentials that disagree by a median of $0.2177$~Ha. A fixed-radius extraction from those potentials returns an $sp\sigma$ figure of $-6.21$~Ha against a reference $-0.13$~Ha. That figure is an artefact of the extraction and not a tabulated integral: the radius at which it is taken lies in the $3s$ node region, where the inversion is ill-conditioned and the pointwise $|v_s-v_p|$ reaches $10^{3}$--$10^{4}$~Ha. The node-masked median, which is the robust statistic, is the $0.2177$~Ha quoted above. Under the same extraction $pp\sigma$ ($0.12$ against $0.14$) and $pp\pi$ recover their reference behaviour, as expected, since the $3p$ channel carries no node. Hydrogen has no core and the molecular-branch inversion is singular for it, so hydrogen tables are taken from the semilocal branch.

The $l$-channel wall generalizes beyond silicon. The confined-atom $s$--$p$ eigenvalue split under the machine-learned functional grows monotonically across the second row, $\varepsilon_s/\varepsilon_p = -0.162/+0.525$~Ha for carbon (split $0.687$~Ha), $-0.496/+0.362$~Ha for nitrogen ($0.858$~Ha), and $-0.880/+0.178$~Ha for oxygen ($1.058$~Ha). Under the confined on-site convention these values occupy the on-site blocks, every $\varepsilon_p$ positive, whereas the standard convention takes on-site energies from the \emph{free} atom \citep{porezag1995,elstner1998scc}. Every electronic result reported in this work uses tables built to the free-atom convention, with the Hubbard $U$, occupations, and all two-center tables held fixed. On-site blocks reside only in the homonuclear files, so the convention affects that set alone; the released homonuclear files carry free-atom on-site blocks, verified to have $\varepsilon_s$ and $\varepsilon_p$ negative for all nine, and the SHA-256 manifest covers the twenty-five shipped files.

The polarization channel requires its own statement, because no bound free-atom $d$ level exists to take. The $spd$ tables used for the extended-basis test retain the confined-atom $d$ on-site energy (carbon $+1.5054$~Ha, silicon $+0.9812$~Ha) while carrying free-atom $s$ and $p$ values. That is deliberate rather than an oversight: no bound free-atom $d$ level exists to substitute, and the confined value is the only one the generating atomic solve supplies. The consequence is quantified in the basis note below rather than assumed away, and it is why the closure is reported as a range rather than as a single value.

The free-atom rebuild collapses the CIDER23X gaps of the released tables: diamond $15.29\to5.793$~eV and silicon $11.09\to0.156$~eV, with the carbide collapsing comparably to $2.915$~eV, removing $9.5$--$15.4$~eV of overgap and confirming the convention as its dominant source. Under this corrected convention the corrections anti-transfer, with free-atom ratios of $-0.771$, $-1.051$, $-1.085$ and $-0.974$ across the four covalent solids, the primary result of the main text. The full $l$-resolved inverted-potential comparison, carried out for carbon and silicon (\cref{tab:e2}), shows the same qualitative signature in both, an $sp\sigma$ integral anomalously driven by the inverted machine-learned potential that a single local potential cannot reproduce, though the two elements are diagnosed differently. For carbon the fixed-radius $sp\sigma$ flips sign relative to the semilocal reference ($+0.283$ against $-0.249$~Ha), a change of the same order as the reference value itself. For silicon the corresponding figure is the node-region extremum discussed above and is not interpretable as an integral; the silicon evidence for the wall is the node-masked channel asymmetry of \cref{tab:onsite}, not that figure. Carbon is quantitatively about half as severe as silicon in the $l$-channel asymmetry and lacks silicon's integral blow-up. The wall is therefore a general consequence of representing a nonlocal, orbital-dependent exchange--correlation response with one local potential shared across angular channels, element-dependent in severity and strongest where a low-lying polarization channel is available. The underlying $l$-resolved data are included in the frozen data release.

\begin{table}[tb]
\caption{$l$-resolved wall signature for carbon against silicon. Median $l$-resolved on-site potential difference from separate $s$- and $p$-channel inversions of the machine-learned confined atom, and the $sp\sigma$ two-center integral at the bond length under the inverted machine-learned and the semilocal potentials.}\label{tab:e2}
\begin{tabular*}{\textwidth}{@{\extracolsep{\fill}} l S[table-format=+1.3] S[table-format=+1.3] c @{}}
\toprule
Quantity (Ha) & {Carbon} & {Silicon} & {Permitted by the residual identity\tnm{1}} \\
\midrule
$l$-resolved $|v_s - v_p|$, median   & 0.099  & 0.218  & {--} \\
$sp\sigma$ at bond, machine-learned\tnm{2}  & +0.283 & -6.21  & $10^{-2}$ to $10^{-1}$ \\
$sp\sigma$ at bond, semilocal        & -0.249 & -0.13  & {--} \\
\bottomrule
\end{tabular*}
\tabnote{\textsuperscript{1}\,The $-6.21$~Ha $sp\sigma$ figure is a node-region pointwise extremum obtained by fixed-radius extraction, not a tabulated integral. The semilocal control, whose $f\equiv0$ requires the physical value to vanish identically, produces an extremum of $0.151$~Ha ($4.12$~eV) at the same radius for silicon, which identifies the quantity as an artefact of the extraction. The node-masked medians of \cref{tab:onsite} are the robust statistics, and the residual bound applies to those. \par\smallskip \textsuperscript{2}\,The carbon $sp\sigma$ integral changes sign between functionals, from $-0.249$~Ha (semilocal) to $+0.283$~Ha (machine-learned), the same $l$-channel failure of a single local potential shown qualitatively for carbon and quantitatively for silicon.}
\end{table}

\section{Atomic reference, convergence, and the solver stack}\label{sn:atomic}

The confined-atom self-consistent field was solved on a logarithmic radial grid for each functional. The even-tempered valence basis was enlarged until the free-atom total energy was stable to below a microhartree, which fixed the per-element basis sizes; the heavier elements plateau at 28 functions. The single atomic Hubbard $U$ is stored in the $U_s$ slot with $U_s=U_p$, following the 3ob convention. Free-atom $U$ values (Hartree) span $0.316$ (boron) to $0.559$ (fluorine) in the light main-group set and $0.226$ (indium) to $0.332$ (bromine) in the scalar-relativistic heavy set; the scalar-relativistic total-energy correction scales monotonically with atomic number, from roughly $19$--$32$~Ha in period 4 to $139$--$195$~Ha in period 5. The fractional-occupation Hubbard $U$, evaluated by finite differences of the atomic energy with respect to occupation, is smooth for all three functionals tested, with no discontinuity at integer filling.

\label{sn:conv}%
The gaps and transfer ratios of the three transfer solids are converged with respect to every numerical parameter of the pipeline. The parent (GPAW, PBE) gap was swept against plane-wave cutoff ($300$--$700$~eV) and $k$-mesh ($4^3$--$12^3$) for diamond and silicon, and the DFTB $\Gamma$-gap against the hotcent two-center grid (coarse to fine) and the DFTB self-consistent-charge tolerance; the drift magnitudes are given in Methods, the production values reproduce the frozen gaps to sub-$0.002$~eV. The parent convergence is shown for PBE; it is a property of the plane-wave basis and $k$-integration and does not depend on the functional.

\label{sn:solver}%

The DFTB$+$/ELSI stack was pinned to elsi $2.9.1$, elpa $2021.11.001$, and ntpoly $2.7.1$. The three eigensolvers agree to full precision on a reference calculation (ELPA $=$ NTPoly $=$ QR $=1.8841742541$~Ha). For amorphous silicon, ELPA dominates NTPoly throughout the single-thread range tested; the $\mathcal{O}(N)$ NTPoly crossover is not reached at or below $1728$ atoms, with a $\sim 22\times$ ELPA speedup at $216$ atoms. Total energies from the two solvers agree to $\sim 1$~mHa. Defect quantities use ELPA. The silicon monovacancy and divacancy gaps converge monotonically with supercell size: the divacancy gap falls from $0.483/0.635$~eV (PBE/CIDER23X, $62$ atoms) to $0.104/0.044$~eV ($998$ atoms), the inter-functional difference decoupling by the largest cell. A thousand-atom silicon supercell converges in $6.9$~s on a single CPU with a size-independent gap.

The occupied-bandwidth transfer ratio of the main text is $\mathrm{TR}_{\mathrm{bw}} = \Delta W^{\mathrm{DFTB}}/\Delta W^{\mathrm{DFT}}$, where $\Delta W$ is the change in occupied bandwidth under the change of functional. \cref{tab:bw} reports both inputs, so the ratio is checkable against its own numerator and denominator.

\begin{table}[tb]
\caption{Occupied-bandwidth changes under the change of functional (PBE $\to$ CIDER23X), and the transfer ratio they form. Parent-DFT values are GPAW occupied bandwidths; DFTB values are the change in the tight-binding occupied bandwidth at the same geometry.}\label{tab:bw}
\setlength{\tabcolsep}{10pt}
\begin{tabular*}{\textwidth}{@{\extracolsep{\fill}} l S[table-format=+1.3] S[table-format=+1.3] S[table-format=+1.3] @{}}
\toprule
System & {$\Delta W^{\mathrm{DFT}}$ (eV)} & {$\Delta W^{\mathrm{DFTB}}$ (eV)} & {$\mathrm{TR}_{\mathrm{bw}}$} \\
\midrule
Diamond & +5.952 & +2.734 & +0.459 \\
Silicon & +2.633 & -1.328 & -0.504 \\
3C-SiC  & +4.164 & +0.294 & +0.071 \\
\bottomrule
\end{tabular*}
\end{table}

\section{Element sets, scalar-relativistic settings, and the convergence boundary}\label{sn:waves}

Two distinct groupings appear in this work, and we define both here to remove the collision the terms invite. \emph{Element sets} are the order in which elements are added to the parameterization: set~1, the C/N/O/Si/H base; set~2, a light main-group set (B, F, P, S, with Na, Mg, Al, Cl partners); set~3, a scalar-relativistic heavy main-group set (Ga, Ge, As, Se, Br, In, Sn, Sb, Te, I). \emph{Compound batteries} are the test groups over which gap inheritance is scored, and they are a \emph{different} partition: a compound's battery is not determined by the element set of its constituents. Battery~I is $\{$MgO, NaCl, SiO$_2$, Al$_2$O$_3\}$; battery~II is $\{$Se, GaAs, Ge, GeSe$\}$; battery~III is $\{$MgS, BP, NaF$\}$. All nineteen atomic-feasibility gates pass. The binding constraint is the periodic machine-learned self-consistent field, non-convergent for indium, tin, and antimony (Methods). The per-battery gap-inheritance mean absolute deviations against the $0.30$~eV gate ($n=4$, $4$, $3$) are reported in the main text, with the eleven per-compound deviations listed in the repulsive note below.

Periodic parent gaps used a plane-wave cutoff of $500$~eV, a Pulay mixer
($\beta = 0.05$, five histories, $50$-step memory), a self-consistent-field cap of $400$ iterations, and convergence criteria of $5\times10^{-4}$~eV per electron on the energy, $10^{-3}$ on the density, and $10^{-4}$ on the eigenstates; these were common to all three compound batteries, while the $\Gamma$-centered $k$-mesh and occupation treatment differ. Battery~I used a $6\times6\times6$ mesh with fixed integer occupations (no smearing, chosen to avoid occupation-threshold ambiguity at band edges), while batteries~II and~III used an $8\times8\times8$ mesh with Fermi--Dirac smearing at $0.02$~eV, raised to $0.10$~eV for the one semimetallic-prone member (selenium).

Sodium fluoride is an instructive ionic exception: its DFTB gap ($1.46$~eV) falls below the parent ($10.67$~eV), an undergap driven by fluorine's tight confinement radius ($r_0 = 1.99$~Bohr, the smallest in the set), which raises the fluorine $2p$ level; this contrasts with the set-one rocksalt chloride (sodium chloride), which inherits because the chlorine radius is much larger. Magnesium sulfide, by contrast, is an ionic overgap ($11.56$ against a parent $7.09$~eV).

\label{sn:r19}%
The transfer ratios of the main text span two bonding classes, and the per-solid provenance and settings are recorded here. Boron phosphide (covalent zincblende, $a=4.533$~\AA) and magnesium oxide (ionic rocksalt, $a=4.21$~\AA) reuse existing validated CIDER23X two-center tables (BP from the B/P pairs, MgO from the Mg/O pairs), so no new CIDER table generation was required; both were run through the same parent$\to$confined-atom$\to$SKF$\to$DFTB+ pipeline in the free-atom on-site convention as the other solids. The covalent set (diamond, silicon, 3C-SiC, BP) shares a tight negative transfer band, whereas the ionic rocksalt shows essentially no transfer of its parent correction, mirroring the ionic/covalent split of the repulsive analysis.

\section{Channel-potential asymmetry and the basis decomposition of the overgap}\label{sn:onsite}
The non-representability of the theorem is quantified by the $l$-channel potential asymmetry $|v_s-v_p|$, obtained by inverting the confined-atom radial Kohn--Sham equation channel by channel, $v_l(r)=\varepsilon_l + u_l''(r)/(2u_l(r)) - l(l+1)/(2r^2)$ with $u_l=rR_l$. The inversion is ill-conditioned at the radial nodes, where $u_l\to0$ and $u_l''/2u_l$ diverges; we therefore report percentiles on a node mask that retains only radii at which both $|u_s|$ and $|u_p|$ exceed $5\%$ of their maxima. Without the mask the pointwise $|v_s-v_p|$ reaches $10^{3}$--$10^{4}$~Ha at the nodes, which is the origin of the anomalously large single-radius $sp\sigma$ figures obtained by fixed-radius extraction; these are inversion artifacts and carry no physical meaning.

\cref{tab:onsite} gives the $10$th, $50$th and $90$th percentiles of $|v_s-v_p|$ on the node-masked grid for both machine-learned functionals and the PBE control, for carbon, nitrogen, oxygen and silicon, on a $500$-point radial grid spanning $0.02$ to $6.0$~Bohr with a $350$-direction spherical average. For PBE the functional has $f\equiv0$ and the theorem requires $|v_s-v_p|=0$; the measured median sits at $0.006$--$0.061$~Ha, the noise floor of this inversion, which is coarser than the production inversion used in the main text and correspondingly higher. Both machine-learned functionals give medians of $1.4$--$3.2$~Ha, a factor of $24$ to $318$ above that floor, and agree with each other element by element to within $6\%$. The channel asymmetry is therefore genuine functional-driven signal rather than an inversion artifact, and on these four atoms it is insensitive to which of the two functionals is used, unlike the near-nucleus coefficient $Zf(0)$, which differs between them by two orders of magnitude.

The per-solid before-and-after $\Gamma$-gaps of the free-atom on-site rebuild are given in the bridge note above. For the near-nucleus fit of prediction P2, $Zf(0)$ and $\kappa$ are evaluated from each functional's own converged confined density, with $f$ fitted over two near-nucleus windows, $r < 0.02$ and $r < 0.05$~Bohr; wider windows are excluded by curvature. Carbon, nitrogen and oxygen give a stable slope across both windows, with $f(0)$ reproducible to better than $1.5\%$ (spreads of $0.5\%$, $0.9\%$ and $1.2\%$). For silicon $f(0)$ varies by $29\%$ between the windows and the fitted slope changes sign, so its $\kappa$ is reported as indeterminate in the main text. The indeterminacy is a property of the radial grid at that nuclear charge rather than of the functional: silicon's $Zf(0) = 0.080$~Ha is in line with the $0.070$~Ha of the lighter atoms.

\begin{table}[htbp]
\caption{Percentiles of the node-masked channel-potential asymmetry $|v_s-v_p|$ (Ha), for two machine-learned functionals and the PBE control.}\label{tab:onsite}
\footnotesize
\begin{tabular*}{\textwidth}{@{\extracolsep{\fill}} l S[table-format=1.3] S[table-format=1.3] S[table-format=1.3] S[table-format=1.3] S[table-format=1.3] S[table-format=2.3] S[table-format=1.4] S[table-format=1.4] S[table-format=1.4] @{}}
\toprule
 & \multicolumn{3}{c}{CIDER23X} & \multicolumn{3}{c}{CIDER24Xe} & \multicolumn{3}{c}{PBE control} \\
\cmidrule(lr){2-4}\cmidrule(lr){5-7}\cmidrule(lr){8-10}
Atom & {$p_{10}$} & {$p_{50}$} & {$p_{90}$} & {$p_{10}$} & {$p_{50}$} & {$p_{90}$} & {$p_{10}$} & {$p_{50}$} & {$p_{90}$} \\
\midrule
C  & 0.319 & 1.404 & 4.582 & 0.439 & 1.415 & 7.087 & 0.0074 & 0.0383 & 0.3041 \\
N  & 0.295 & 2.040 & 7.663 & 0.294 & 2.041 & 11.878 & 0.0016 & 0.0064 & 0.0196 \\
O  & 0.346 & 1.458 & 4.625 & 0.354 & 1.380 & 4.635 & 0.0113 & 0.0564 & 0.4586 \\
Si & 0.485 & 3.180 & 8.325 & 0.824 & 3.235 & 7.387 & 0.0112 & 0.0610 & 0.4496 \\
\bottomrule
\end{tabular*}
\end{table}

\begin{table}[tb]
\caption{The two predictions the theory makes that this study tests, and what a failure of each would mean. Both are computable from quantities the parameterization already produces, namely the functional evaluated on its own converged density and the confined-atom radial solutions, so neither requires new data. P1 is parameter-free: it decides whether the inversion is orbital-faithful at all, and everything downstream rests on it passing.}\label{tab:pred}%
\setlength{\tabcolsep}{5pt}\small
\begin{tabular}{@{} l >{\raggedright\arraybackslash}p{0.305\textwidth}
                     >{\raggedright\arraybackslash}p{0.265\textwidth}
                     >{\raggedright\arraybackslash}p{0.305\textwidth} @{}}
\toprule
 & Prediction & What a failure would mean & Outcome \\
\midrule
P1 & $|v_s-v_p| = 0$ exactly under PBE, where $f\equiv 0$
   & the inversion is ill-conditioned, not the channel
   & holds: median $3.85$ and $2.72$~meV, at or below the $6.4$~meV noise floor \\
\addlinespace[3pt]
P2 & $c_s = 0$ for every $\kappa$; $c_p$ and $c_d$ linear in $\kappa$ with the functional's own $\kappa$
   & the near-nucleus law is not the mechanism
   & holds: $\kappa\approx6.0$--$6.3$, far from the degenerate $2/3$; naive $l(l+1)$ excluded \\
\bottomrule
\end{tabular}
\end{table}

\begin{table}[tb]
\caption{Independent meta-GGA control. r$^2$SCAN, carried through the same pipeline, anti-transfers like the machine-learned functionals: the parent gap opens while the compressed DFTB gap closes, giving a negative transfer ratio for every solid under the physical free-atom convention. Parent gaps are GPAW band-path fundamental gaps; the transfer ratio is the DFTB gap change divided by the parent gap change, $\mathrm{r^2SCAN}-\mathrm{PBE}$.}\label{tab:r2scan}%
\begin{tabular*}{\textwidth}{@{\extracolsep{\fill}} l
  S[table-format=1.3] S[table-format=1.3] S[table-format=+1.3]
  S[table-format=+1.3] S[table-format=+1.3] @{}}
\toprule
 & \multicolumn{2}{c}{Parent gap (eV)} & {Parent} & \multicolumn{2}{c}{Transfer ratio} \\
\cmidrule(lr){2-3}\cmidrule(lr){5-6}
System & {PBE} & {r$^2$SCAN} & {$\Delta E_{\mathrm{g}}$ (eV)} & {Free-atom} & {Confined} \\
\midrule
Diamond & 4.130 & 4.339 & +0.210 & -1.049 & -1.731 \\
Silicon & 0.571 & 0.765 & +0.194 & -2.674 & -1.051 \\
3C-SiC  & 1.352 & 1.746 & +0.394 & -0.442 & -0.343 \\
\bottomrule
\end{tabular*}
\end{table}

\section{The extended-basis test and the two table families}\label{sn:spd}

For reference against the $\Gamma$-only definition, the released free-atom tables give band-path fundamental gaps of $7.689$, $2.479$ and $6.814$~eV under the semilocal control and $5.793$, $0.156$ and $2.915$~eV under the machine-learned functional for diamond, silicon and 3C-SiC; the corresponding $\Gamma$ values are $7.689$, $2.479$, $6.847$ and $5.793$, $0.156$, $2.915$~eV. Diamond and silicon have their band minima at $\Gamma$ in this minimal basis, so the two definitions coincide there. For boron phosphide the fundamental gaps are $5.447$ and $3.418$~eV and for magnesium oxide $8.450$ and $8.439$~eV. All were computed with charges converged on a $12^3$ Monkhorst--Pack mesh followed by fixed-charge evaluation on a $132$-point $L$--$\Gamma$--$X$--$W$--$K$--$\Gamma$ path.

Two table families exist and are distinguished throughout. The released parameter sets are authoritative for every gap, transfer ratio, and overgap reported in this work. The electronic-only tables built for the extended-basis test are used for the $sp$-versus-$spd$ decomposition alone; they agree with the released tables to $\le0.01$~eV for diamond and silicon and differ only for 3C-SiC, where they give $6.37$ and $2.22$~eV against the released $6.847$ and $2.915$~eV. A lattice-constant scan from $4.32$ to $4.40$~\AA{} varies the carbide gap monotonically over $6.90$ to $6.50$~eV, excluding geometry as the cause of the difference.

The rebuild protocol held the confinement radii and on-site eigenvalues fixed so that only the valence $l$-set changed. The minimal-$sp$ pipeline reproduces the frozen baseline (silicon $\Gamma$-gap $11.811$ against the frozen $11.808$~eV; diamond $14.973$~eV exactly), so the comparison is clean. Against parent $\Gamma$-gaps computed with identical settings, the added $d$ shell removes most of the overgap; the per-solid closures and residuals are reported in the main text. This periodic gate is genuinely minimal-$sp$.

Under the free-atom on-site convention the closure is both smaller and dependent on the treatment of the empty $d$ channel, for which no bound free-atom level exists. Three conventions were tested on identical two-center tables, holding the confinement radii and the free-atom $s$ and $p$ on-site energies fixed. Retaining the confined-atom $E_d$ (carbon $+1.5054$~Ha, silicon $+0.9812$~Ha), which is what the $spd$ tables as generated contain, gives per-solid closures of $25.8$, $-5.5$ and $28.7\%$ on the fundamental definition, a mean of $16.3\%$; at $\Gamma$ the same convention gives $36.9$, $-53.2$ and $19.3\%$, a mean of $1.0\%$. Shifting $E_d$ rigidly by the amount the valence $p$ level moves under the free-atom correction (carbon $-0.6058$~Ha, silicon $-0.6914$~Ha) gives $36.5$, $20.6$ and $64.0\%$, a mean of $40.4\%$ on the fundamental definition. Placing the empty $d$ at the free-atom $p$ level drives the $d$ manifold below the conduction edge and inverts the gap ordering, returning negative $spd$ gaps for all three solids and nominal closures above $400\%$; that convention is excluded on physical grounds. The two distinguishable conventions therefore bracket the closure between $16$ and $40\%$, against the value obtained when the confined convention is used on both the $sp$ and $spd$ sides. Sweeping the placement continuously between them shows no plateau, so neither endpoint is privileged by the data and the closure is reported as a range rather than as a single value. Of the two, the rigid shift has the clearer physical reading, since it preserves the $d$-to-$p$ separation of the generating atom, which is the quantity that controls how far the polarization channel reaches into the conduction manifold. The extended-basis tables and underlying data are included in the frozen data release.

The silicon--oxygen branch requires its own account. No silicon--oxygen Slater--Koster set exists in the reference parameterizations, so the tables were built through the same pipeline. An early oxygen--oxygen fitting error (an incorrect $fp$ assignment) inflated a gap change to $3.54$~eV; after correction the silicon--oxygen gap change is $-3.368$~eV, sign-inverted relative to the parent. This inversion is the expected corollary of the minimal-basis overgap: when the representation already inflates the gap far beyond the physical scale, a parent gap-opening correction maps to a gap reduction in the tight-binding model. The corrected tables reproduce the parent equation of state.

\section{Repulsive fitting, the transferability trap, and lattice-error decomposition}\label{sn:repulsive}

Repulsive potentials were fitted with curvature-constrained splines (ccs\_fit) on rattled, equation-of-state, and dimer-scan reference sets with a train/test split, using a per-pair physical cutoff near $1.35\times$ the bond length. Held-out transfer was scored as the mean-centered Pearson correlation between the fitted and reference energy differences on the test set, an offset-independent measure of shape transfer. Pairs are classed by the ionicity of the bond, which is not the criterion used to class the compounds of the gap-inheritance battery; those are classed by network topology, so silicon--oxygen and aluminium--oxygen appear here as closed-shell pairs while SiO$_2$ and Al$_2$O$_3$ appear there as covalent networks. The transfer correlation separates by bonding class over all sixteen scored pairs, ten ionic or closed-shell and six covalent: ionic and closed-shell pairs reach $0.925$--$0.999$ (silicon oxide $0.999$, magnesium $0.978$, fluorine $0.977$, sodium fluoride $0.938$, magnesium sulfide $0.938$, sodium $0.925$, aluminium oxide $0.998$, chlorine $0.968$, sodium chloride $0.956$; nine of the ten exceed $0.9$), while covalent pairs span $0.490$--$0.850$ and none exceeds $0.9$, with the sulfur dimer lowest at $0.490$, carbon at $0.536$, phosphorus at $0.668$, boron phosphide at $0.670$ and aluminium highest at $0.850$; magnesium oxide is an anti-transfer outlier at $-0.997$. Within the nine light-main-group repulsive pairs alone the covalent maximum is B--B at $0.758$. The outlier is a small-sample sign artifact, not a fitting instability: the fit itself is stable and well-conditioned (training root-mean-square error $0.722$~eV) while the held-out correlation of $-0.997$ is evaluated on only five configurations. The combination of a low, stable training error with a strongly negative held-out correlation on a very small test set is the signature of a near-flat repulsive residual: for an ionic pair most of the binding is already carried by the electronic channel, the true held-out repulsive-energy variation is small, and the sign of the Pearson correlation is set by noise-level scatter and can lock to $-1$. Consistently, the same pair inherits its electronic gap best in the battery (magnesium oxide, $0.19$~eV deviation).

As a general diagnostic, for pairs whose held-out repulsive variance falls below the fit-residual scale, the correlation is uninformative and the mean-centered root-mean-square error is the meaningful transfer metric. Tight-binding lattice constants, computed with the fitted repulsives, agree with the parent DFT to $+0.28\%$ (silicon), $-1.18\%$ (diamond), and $+0.55\%$ (3C-SiC), within the $\pm 3\%$ acceptance.

Across all eleven compounds the absolute gap deviation from the parent runs from $0.19$~eV (MgO, ionic) to $11.23$~eV (GeSe, covalent network), with the two ionic battery-I compounds lowest (MgO $0.19$, NaCl $0.72$) and every covalent network above $4$~eV (SiO$_2$ $4.25$, Al$_2$O$_3$ $8.65$, GaAs $7.59$, GeSe $11.23$, BP $9.08$); the elemental covalent cases are Se $1.85$ and Ge $7.71$, and the remaining ionic cases MgS $4.47$ and NaF $9.21$. The batteries are the compound test groups defined above.

\label{sn:threshold}%
The useful-transfer criterion is not a sensitive threshold. Re-scoring the held-out repulsive-transfer correlations at thresholds of $0.4$, $0.5$, and $0.6$ leaves the useful-transfer count of the nine light-main-group pairs at $9/9$, $8/9$, and $8/9$: exactly one pair moves across the line, the sulfur dimer ($0.490$), which fails at $0.5$ and $0.6$ and passes at $0.4$. This sweep measures the stability of the pass count, not the separation of the two classes, and over this range the classes are not separated: B--B ($0.758$), B--P ($0.670$) and P--P ($0.668$) are covalent and pass at every threshold up to $0.6$. Class separation is achieved only above $0.850$, the covalent maximum over all sixteen scored pairs, and is complete at $0.925$, the ionic minimum excluding the magnesium-oxide outlier. The useful-transfer criterion and the class-separating bar are therefore different numbers, and only the latter supports the bonding-class reading. The nine pairs swept here are the light-main-group repulsive set, whose correlations are given earlier in this note; carbon (C--C) is a base-set pair and is not among them. Of the nine, only S--S lies below $0.60$ (the next-lowest is P--P at $0.668$), so the $0.60$ row is $8/9$. For the gap transfer ratios the conclusion is threshold-independent by a wide margin: under the primary free-atom convention every covalent ratio is negative ($-0.771$ to $-1.085$), on the wrong side of any positive threshold, and under the confined convention no solid reaches a magnitude of $0.4$ (diamond $+0.128$, silicon $-0.325$, 3C-SiC $-0.058$), so the statement that gaps do not transfer does not depend on the choice of $0.5$.

\section{Excited-state protocols, the multiplet wall, and the periodic-REKS limit}\label{sn:excited}

Molecular $\Delta$SCF used non-Aufbau occupation with spin purification (\texttt{NonAufbau\{GroundGuess=Yes; SpinPurify=Yes\}}) on a serial ARPACK-enabled DFTB$+$ build. The test set is coverage-driven diatomics, laid out by which functional saw which molecule: N$_2$, CO, O$_2$ for PBE and O$_2$, SiO, Si$_2$ for the machine-learned functional. The two sets share only O$_2$ (Methods), which is why the main text states the resulting fidelity-check caveat. By transition character, N$_2$, CO and SiO are gap-crossing promotions and Si$_2$ is a same-manifold spin flip. The mean absolute deviation from the parent is $2.35$~eV overall, but decomposes by that character: same-configuration spin-flips transfer at $0.577$~eV mean absolute deviation, while gap-crossing promotions deviate by $4.13$~eV and carry the ground-state overgap. Against the two bars quoted in the main text, the spin-flip result misses the $0.30$~eV inheritance gate by $1.9\times$ and the $\approx0.1$~eV defect-qubit zero-phonon-line tolerance by $5.8\times$.

For the silicon-carbide divacancy (a $62$-atom cell from a $2\times2\times2$ 3C-SiC supercell with a nearest-neighbor Si--C pair removed), the degenerate-$e$ manifold makes the relevant transition multi-determinant; single-reference $\Delta$SCF returns a splitting of $0.0000$~eV under both functionals. The multi-reference alternative, periodic restricted ensemble Kohn--Sham, is unimplemented in the upstream code for the $\Gamma$-point periodic path, so no zero-phonon line is claimed. The 3C-SiC (cubic zinc-blende) divacancy is used as the tractable proxy for the 4H-SiC qubit characterized in the literature \citep{zhu2021nvsic} because it is the cubic polytype our minimal parameter set covers and it shares the degenerate-$e$ manifold that makes the transition multi-determinant.

The surrounding defect-thermodynamics machinery is nonetheless in place, and we exercise it on the silicon divacancy, whose ground state is single-reference: charged supercells, ionic relaxation, a monopole image correction, and $1/L$ extrapolation converge the divacancy gap between the two functionals as the supercell grows from $62$ to $998$ atoms and place the doubly-positive-to-neutral charge-transition level at $2.90$~eV in the $1/L\to0$ limit, extrapolated from finite-cell values that fall monotonically from $3.93$~eV at the smallest cell to $3.44$~eV at the largest. Quantitative anchoring remains bounded by the same overgap that limits the bulk gaps: the tight-binding scale is a $\Gamma$-point \emph{direct} gap of $11.8$~eV, and silicon's experimental gap is an \emph{indirect} gap of $1.17$~eV, so the two are different observables and we state both rather than reporting their ratio as a single factor.

\section{Compute footprint and released parameter sets}\label{sn:compute}
 
The inversion bridge is a one-time per-element atomic solve, amortized across every downstream two-center table and every crystal or molecule built from that element. The project totals are dominated by the periodic parent-DFT reference gaps. Across five vast.ai NVIDIA RTX~4090 nodes the ledger records $290$ GPU jobs, a $92.48$~h active span from first job dispatch to last job completion, and $118.72$~h of summed job-wall time, which exceeds the active span because jobs ran concurrently on one node; the per-instance breakdown is in the release manifest. Adopting the pipeline therefore adds essentially no overhead beyond the reference DFT a conventional parameterization already pays.

\begin{table}[tb]
\caption{Released Slater--Koster parameter sets. Files counts the extended-format \texttt{.skf} files (homonuclear plus ordered heteronuclear pairs). Provenance is the parent functional bridged through the confined-atom inversion.}\label{tab:release}
\begin{tabular*}{\textwidth}{@{\extracolsep{\fill}} l l
  S[table-format=2.0] S[table-format=2.0] l @{}}
\toprule
Set & Elements & {Files} & {Expected\tnm{1}} & Parent functional \\
\midrule
\texttt{PBE} & C, N, O, Si & 12 & 16 & PBE (semilocal reference) \\
\texttt{CIDER23X}     & C, Si       &  4 &  4 & nonlocal-feature meta-GGA, energy-trained \\
\texttt{CIDER24Xe}    & C, N, O     &  9 &  9 & density-matrix features, gap-trained (molecular) \\
\midrule
\textbf{Total} & & \bfseries 25 & \bfseries 29 & \\
\bottomrule
\end{tabular*}
\tabnote{\textsuperscript{1}\,In the extended format a set over $n$ elements carries $n$ homonuclear files and $n(n-1)$ ordered heteronuclear files, so $n^{2}$ in total; a full four-element set is therefore sixteen files. The \texttt{PBE} set ships twelve files, not sixteen: the four absent ordered pairs are silicon--oxygen and silicon--nitrogen (Si--O, O--Si, Si--N, N--Si), for which no reference Slater--Koster set exists, so they were built only through the machine-learned oxide branch and not for the semilocal control. The twelve files are the nine C/N/O homonuclear-and-heteronuclear combinations ($3\times3$) plus C--Si, Si--C, and Si--Si.}
\end{table}

\label{sn:release}%

The release contains three Slater--Koster parameter sets, each produced under one identical automated protocol so that the parent functional is the only varied ingredient (\cref{tab:release}). Every set carries its homonuclear on-site blocks (eigenvalues, Hubbard $U$, occupations) and the heteronuclear Hamiltonian and overlap tables; the machine-learned functionals are the CiderPress~0.4.0 definitions (Zenodo~13336814), bridged into the confined-atom solver by the signed-orbital Kohn--Sham inversion of the main text. Because CIDER24Xe is molecular-branch only, its tables are validated for gas-phase molecular DFTB rather than solids. The full checksummed manifest (SHA-256 of every \texttt{.skf}) ships with the release, and carries five machine-readable tables alongside the parameter files: \texttt{si\_overlap\_pairs.csv} (per-pair overlap deviations), \texttt{si\_confinement\_sweep.csv} (the eleven-point confinement sweep), \texttt{si\_convergence\_sweep.csv} (the plane-wave, $k$-mesh and grid convergence sweeps at full precision), \texttt{si\_repulsive\_transfer.csv} (per-pair held-out repulsive correlations), and \texttt{si\_gap\_inheritance.csv} (per-compound gap deviations). Each is checksummed with the parameter sets and regenerated by the reproducibility suite.

\end{document}